\documentclass[aps,prb,twocolumn,superscriptaddress]{revtex4}
\pdfoutput=1
\usepackage{amsmath}
\usepackage{amssymb}
\usepackage{bm}
\usepackage{bbm}
\usepackage{graphicx}
\usepackage{times}
\usepackage{color} %for change tracking
\usepackage{ulem}
\usepackage{subfigure}

\newcommand{\ket}[1]{|#1\rangle}
\newcommand{\bra}[1]{\langle#1}

\newcommand{\eq}{\begin{equation}}
\newcommand{\fine}{\end{equation}}

\usepackage{cancel}
\usepackage{soul}

\begin{document}

\title{Photonic polarization gears for ultra-sensitive angular measurements}

\author{Vincenzo D'Ambrosio}
\affiliation{Dipartimento di Fisica, Sapienza Universit\`a di Roma, piazzale Aldo Moro 5, 00185 Roma, Italy}
\author{Nicol\`o Spagnolo}
\affiliation{Dipartimento di Fisica, Sapienza Universit\`a di Roma, piazzale Aldo Moro 5, 00185 Roma, Italy}
\author{Lorenzo Del Re}
\affiliation{Dipartimento di Fisica, Sapienza Universit\`a di Roma, piazzale Aldo Moro 5, 00185 Roma, Italy}
\author{Sergei Slussarenko}
\affiliation{Dipartimento di Fisica, Universit\`a di Napoli Federico II, Compl. Univ. di Monte S. Angelo, Napoli, Italy}
\author{Ying Li}
\affiliation{Centre for Quantum Technologies, National University of Singapore, Singapore 117543}
\author{Leong Chuan Kwek}
\affiliation{Centre for Quantum Technologies, National University of Singapore, Singapore 117543}
\affiliation{Institute of Advanced Studies (IAS), Nanyang Technological University, Singapore 639673}
\affiliation{National Institute of Education, Nanyang Technological University, Singapore 637616}
\author{Lorenzo Marrucci}
\affiliation{Dipartimento di Fisica, Universit\`a di Napoli Federico II, Compl. Univ. di Monte S. Angelo, Napoli, Italy}
\affiliation{CNR-SPIN, Complesso Universitario di Monte S. Angelo, Napoli, Italy}
\author{Stephen P. Walborn}
\affiliation{Instituto de F\'{i}sica, Universidade Federal do Rio de Janeiro, Rio de Janeiro, RJ 21941-972, Brazil}
\author{Leandro Aolita}
\affiliation{ICFO-Institut de Ci\`encies Fot\`oniques, Parc Mediterrani
 de la Tecnologia, 08860 Castelldefels (Barcelona), Spain}
\affiliation{Dahlem Center for Complex Quantum Systems, Freie Universit\"{a}t Berlin, Berlin, Germany}
\author{Fabio Sciarrino}
\email{fabio.sciarrino@uniroma1.it}
\affiliation{Dipartimento di Fisica, Sapienza Universit\`a di Roma, piazzale Aldo Moro 5, 00185 Roma, Italy}
\affiliation{Istituto Nazionale di Ottica (INO-CNR), Largo E. Fermi 6, I-50125 Firenze, Italy}

%%%%%%%%%%%%%%%%%%%%%%%%%%%%%%%%%%%%%%%%%%%%%%%%%%
\maketitle
%%%%%%%%%%%%%%%%%%%%%%%%%%%%%%%%%%%%%%%%%%%%%%%%%%

{\bf Quantum metrology bears a great promise in enhancing measurement precision, but is unlikely to become practical in the near future. Its concepts can nevertheless inspire classical or hybrid methods of immediate value. Here, we demonstrate NOON-like photonic states of $m$ quanta of angular momentum up to $m=100$, in a setup that acts as a ``photonic gear'', converting, for each photon, a mechanical rotation of an angle $\theta$ into an amplified rotation of the optical polarization by $m\theta$, corresponding to a ``super-resolving'' Malus' law. We show that this effect leads to single-photon angular measurements with the same precision of polarization-only quantum strategies with $m$ photons, but robust to photon losses. Moreover, we combine the gear effect with the quantum enhancement due to entanglement, thus exploiting the advantages of both approaches. The high ``gear ratio" $m$ boosts the current state-of-the-art of optical non-contact angular measurements by almost two orders of magnitude.}

The precise estimation of a physical quantity is a relevant problem in many research areas. Classical estimation theory asserts that by repeating an experiment $N$ times, the precision of a measurement, defined by the inverse statistical error of its outcome, can be increased at most by a factor of $\sqrt{N}$. In quantum physics, this scaling is known as the standard quantum or shot-noise limit, and it holds for all measurement procedures that do not exploit quantum effects such as entanglement. Remarkably, using certain $N$-particle entangled states it could be possible to attain a precision that scales as $N$. This is known as the Heisenberg limit, and is the ultimate bound set by the laws of quantum mechanics \cite{GiovannettiReview}. Proof-of-principle demonstrations of these quantum-metrology concepts have been given in recent experiments of optical-phase estimation, magnetic-field sensing and frequency spectroscopy \cite{Mitchell04,Walther04,Nagata07,Afek10,Spag12,Leibfried05,Roos06,Jones09}. In photonic approaches, the optimal measurement strategy typically relies on the preparation of ``NOON" states \cite{Lee02}, in which all $N$ photons propagate in one arm or the other of an interferometer. However, the experimental preparation of NOON states with large $N$ is extremely challenging, and to date only $N=3$, $N=4$, and $N=5$ photonic NOON states have been reported \cite{Mitchell04,Walther04,Nagata07,Afek10}.  Moreover, as $N$ grows, $N$-photon entangled states become increasingly sensitive to losses, as the loss of a single photon is enough to destroy all the phase information \cite{Dorner}. It has been proved that, in the presence of losses or other types of noise, no two-mode quantum state can beat the standard limit by more than just a constant factor in the limit of large $N$ \cite{kolodynski10,escher11,Knysh,Demk12}.

%%%%% FIGURE 1 %%%%%
%%%%%%%%%%%%%%%%%%%%%%%%%%%%%%%%%%%%%%%%%%%%%%%%%%
\begin{figure*}[t]
\includegraphics[width=0.95\textwidth]{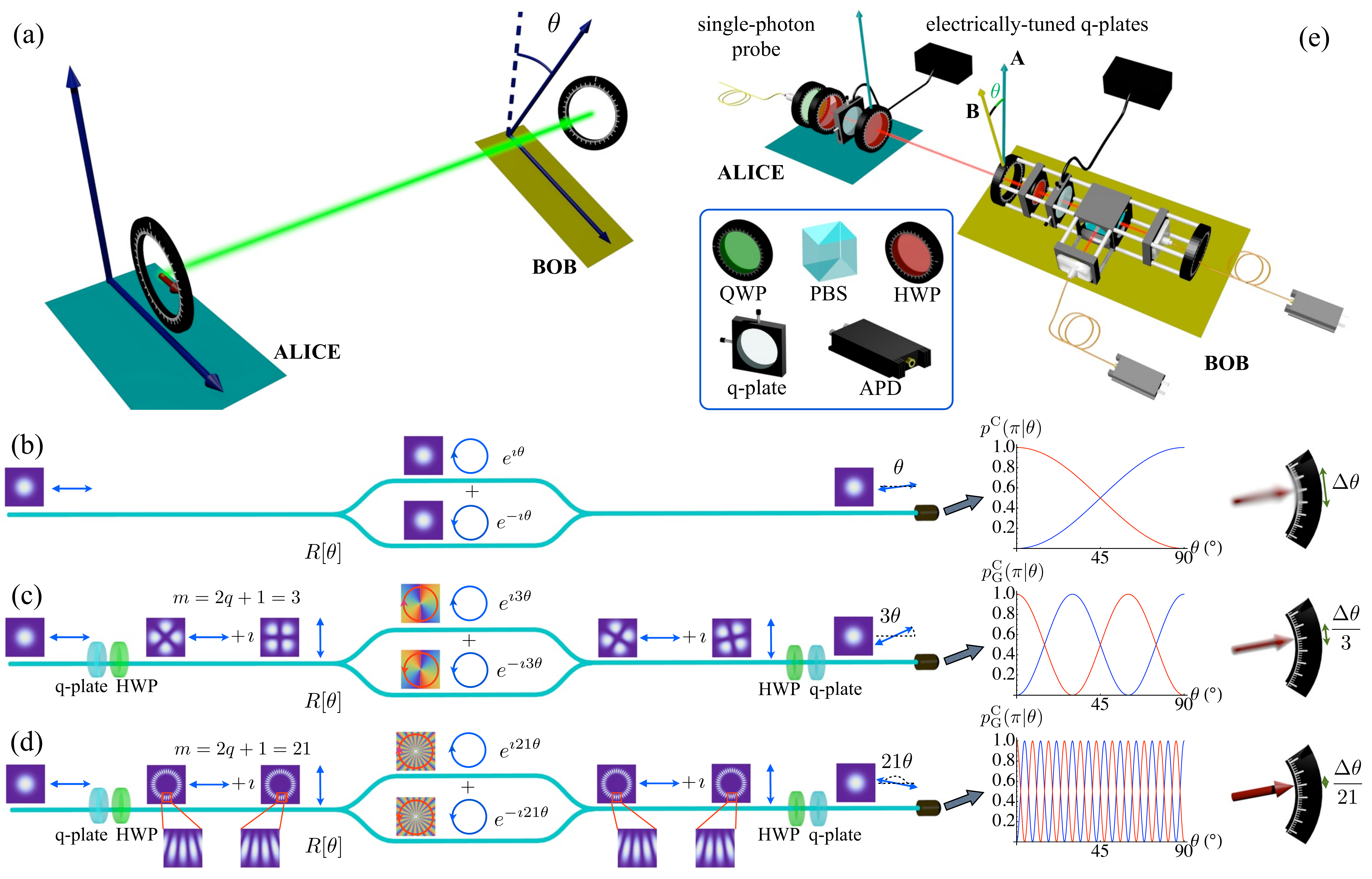}
\caption{\textbf{Photonic gear concept.} (a)  A sender Alice prepares and sends to a receiver Bob photonic probes to measure the relative angle $\theta$ between their reference frames. (b-d) Equivalent interferometric scheme. The action of the physical rotation can be schematically represented as an interferometer, where the two arms correspond to the right- and left-circular components of the photon. (b) Polarization-only states are used. The physical rotation introduces a relative phase between the right- and left-circular components of the photon, corresponding to a rotation of the final photon polarization by the same angle $\theta$. The measurement is repeated $\nu$ times, and polarization fringes $p^{\mathrm{C}}(\pi \vert \theta)= \cos^2\theta$ are recorded (with $\pi=H,V$), from which the angle $\theta$ is retrieved with a statistical error $\Delta \theta$ (represented as a blurred arrow pointing to a goniometer). (c-d) Hybrid SAM-OAM photon states are generated by exploiting $q$-plates [(c) $q=1$, (d) $q=10$] and are used to estimate the angle $\theta$. The physical rotation introduces a relative phase between the two components which varies $m=2q+1$ times faster than the polarization-only case, so that the output photon polarization rotates $m$ times faster (photonic gear effect). The recorded polarization fringes $p^{\mathrm{C}}_{\mathrm{G}}(\pi \vert \theta)$ after decoding with a second $q$-plate now present a periodicity $\propto 1/m$, leading to an improved angular sensitivity $\Delta \theta/m$. The intensity (squared blue contour plots) and phase patterns (squared contour plots in false colors) of the linear and circular polarization components of the employed SAM-OAM states are also shown. (e) Experimental setup. In the single-photon regime, Alice uses photons generated by a parametric down-conversion heralded source. In the classical regime, Alice uses coherent laser pulses. The quantum regime, in turn, uses entangled photons and is described in Fig. \ref{fig:entanglement}. Bob's detection apparatus is mounted in a compact and robust stage which can be freely rotated around the light propagation axis \protect\cite{D'Ambrosio12}. Legend: QWP - quarter-wave plate, HWP - half-wave plate, PBS - polarizing beam-splitter, APD - fiber-coupled single-photon detector.}
\label{QPlate}
\end{figure*}
%%%%%%%%%%%%%%%%%%%%%%%%%%%%%%%%%%%%%%%%%%%%%%%%%%

Here we demonstrate the preparation of single-photon NOON-like quantum states that are superpositions of eigenstates of light with opposite total angular momentum quantum numbers. By total angular momentum, we refer here to the sum of the spin-like angular momentum (SAM), associated with left and right circular polarization states, and of the orbital angular momentum (OAM) that characterizes helical modes of light \cite{Moli07,Fran08}. While SAM can have only two values $\pm\hbar$, where $\hbar$ is the reduced Planck constant, the OAM per photon can generally be given by $l\hbar$, where $l$ is an arbitrarily large positive or negative integer. The SAM-OAM superposition states that we generate involve $m=l+1$ quanta of angular momentum $\hbar$, oriented either parallel or anti-parallel to the propagation axis of each photon. By exploiting such states in the single-photon regime, mechanical rotations can be measured with a precision scaling as $m$ times the square root of the number of probes used. Such enhanced sensitivity can be seen as resulting from a ``super-resolving'' interference between the two $m$-quanta angular momentum orientations appearing in the superposition, analogously to the two arms of the NOON-state interferometers. Notably, in this regime every photon is disentangled from all others and hence the loss of a photon does not affect the overall phase coherence, making the scheme loss-robust. Moreover, the experimental state production and detection are exponentially more efficient than for $N$-photon entangled states. We notice that algorithms for photonic phase estimation without multi-photon entanglement have already been realized \cite{Higgins07}. However, these rely on repeated applications of the unknown phase-shift to be measured, and remotely coordinated (between Alice and Bob) adaptive measurements. Furthermore, these approaches are exponentially (in the number of applications) sensitive to losses. Although quantum-inspired, our approach is essentially classical, because the enhancement does not come from quantum entanglement but results instead from the rotational sensitivity of large angular momentum eigenmodes. In fact, our photonic gears can operate also in the fully classical regime, as described by coherent states.
The sensitivity enhancement for a given mechanical rotation is obtained in the form of an ``amplified'' rotation of the uniform polarization state of the light, that is the mentioned ``polarization gears'' effect.  Rotation sensors based on OAM have been reported before \cite{jha,zeilinger,padgett90s,barnett,leach}, but our approach is qualitatively different from all other OAM-related proposals in the fact that we use SAM-OAM combined states that allow us to ``read'' the rotation by a simple polarization measurement, thus without introducing the large photon losses arising from diffraction or transmission in the angular masks usually needed to read the OAM state. When inserting the photonic gears between polarizers, we observe a ``super-resolving'' Malus' law, reminiscent of the polarization-correlations visible with multi-photon quantum states \cite{Mitchell04,Walther04}, but which appears in both the classical light or single photon regimes.

In this work, we test our photonic gears in three different regimes: (i) classical intense laser light; (ii) single-photon regime, that we adopt to quantitatively compare the achieved angular sensitivity with the shot-noise and Heisenberg limits; (iii) quantum regime of entangled photons, in which we demonstrate that the photonic gears can be combined with quantum correlations, leading to different kinds of ``super-resolving'' rotational correlations between the two measurement stages receiving the two photons. In particular, we produce a quantum state that is metrologically equivalent to a NOON state, leading to a hybrid quantum-classical enhancement of the angular sensitivity. The precision attained in this case scales as $mN$ times the square root of the number of probes used, the $m$ originating from the gear ratio and the $N$ (instead of the classical $\sqrt{N}$) from quantum entanglement. We perform a proof-of-principle demonstration with $N=2$ and total angular momentum up to 18.

\section*{Results}
{\bf Photonic gear concept}. The key element for our photon state manipulations is the $q$-plate, a novel liquid crystal device that efficiently maps pure polarization states into hybrid SAM-OAM states and vice versa \cite{Marr06,Naga09,Marr11} (see Methods and Supplementary Figure S1 for details). For a linearly polarized input, the photonic states generated are superpositions of $m=2q\pm1$ quanta in opposite total angular momentum eigenmodes, where $q$ is an integer or semi-integer ``topological charge'' characterizing the device. Previous achievements were limited to $q$-plates with low $q$ (up to 3) \cite{Slus11}. In particular, rotational-invariant states with $m=0$ ($q=1/2$) were recently used to demonstrate alignment-free quantum communication \cite{D'Ambrosio12}. Here, we introduce a new family of devices with $q$ ranging up to 50, producing angular momentum values as large as $m=101$. These photonic states can be also classically visualized as space-variant polarization states \cite{Carda12}. When these SAM-OAM superposition states are passed through a second $q$-plate, they are converted back into pure polarization states with zero OAM and a uniform polarization. However, a relative rotation of the transmitting and reading stages by a given angle $\theta$ is converted into a rotation of the light optical polarization by the angle $m\theta$, which in our case can be as high as $101\theta$. It is this ``gear ratio'' $m$ that gives rise to the angular sensitivity enhancement.

In the following, we explain in greater detail the photonic gears concept by adopting a quantum language, with the purpose of comparing our sensitivity enhancement with the shot-noise and Heisenberg limits and to allow an easier generalization to the hybrid case in which there is both a classical and a quantum effect. Let us then consider the scenario where a sender Alice and a receiver Bob wish to measure a relative misalignment angle $\theta$ between their reference frames around the optical axis [see Fig. \ref{QPlate} (a)]. A classical strategy for this task consists of Alice sending $N$ photons [see Fig. \ref{QPlate} (b)], each one in state $\ket{\Psi^\mathrm{C}}\doteq\ket{1}_H\equiv\frac{1}{\sqrt{2}}(\vert 1\rangle_R+\vert 1\rangle_L)$, where $\vert n \rangle_x$ denotes a state of $n$ photons in mode $x$, with $x=H$, $R$, or $L$, representing the horizontal-linear, right- and left-circular polarization modes, respectively. All modes in the vacuum state are omitted for brevity. Bob fixes a polarizer in the $H$ direction in his coordinate system, where the misalignment corresponds to a rotation by $-\theta$ of the photons' state. In turn, the $L$ and $R$ polarization states are eigenstates of rotation, so that in Bob's frame $\ket{\Psi^\mathrm{C}}$ becomes $\ket{\Psi^\mathrm{C}(\theta)}=\frac{1}{\sqrt{2}}(e^{i\theta}\vert 1\rangle_R+e^{-i\theta}\vert 1\rangle_L)$. The conditional  probability that he detects a photon in the $H$-polarization (of his reference frame) given that the phase is $\theta$ is given by Malus' law: $p^\mathrm{C}(H|\theta)=\cos^2\theta$. By measuring this probability, Alice and Bob can estimate $\theta$. To strengthen their statistics, they repeat the procedure $\nu$ times, consuming a total of $\nu\times N$ photons, and average all the outcomes.
Their final statistical error is bounded as
\begin{equation}
\label{DeltaC}
\Delta\theta^\mathrm{C} \geq \Big[2\sqrt{\nu N}\Big]^{-1}.
\end{equation}
The right-hand side is the standard quantum limit, and can always be reached in the asymptotic limit of large $\nu N$ \cite{GiovannettiReview}. Our error estimators $\Delta\theta$  are standard root-mean-squared variances. In general, for phases, a cyclic error cost-function would be more appropriate, as for instance the Holevo variance \cite{Holevo}. However, both types of variances coincide in the small-error limit, so for our purposes the standard variance is adequate.

Using quantum resources, the optimal strategy consists of Alice sending $\nu$ probes, each one composed of the $N$-photon entangled NOON state $\ket{\Psi^\mathrm{Q}}=\frac{1}{\sqrt{2}}(\vert N\rangle_R+\vert N\rangle_L)$. In Bob's frame, this state is expressed as $\ket{\Psi^\mathrm{Q}(\theta)}=\frac{1}{\sqrt{2}}(e^{i N\theta}\vert N\rangle_R+e^{-i N\theta}\vert N\rangle_L)$. The conditional probability that he detects the unrotated state $\ket{\Psi^\mathrm{Q}}$ is $p^\mathrm{Q}(\Psi^{\mathrm{Q}}|\theta)\doteq|\bra{\Psi^\mathrm{Q}}\ket{\Psi^\mathrm{Q}(\theta)}|^2=\cos^2(N\theta)$, which resolves values of $\theta$ that are $N$ times smaller than $p^\mathrm{C}(H|\theta)$. Their uncertainty is then bounded as 
\begin{equation}
\label{DeltaQ}
\Delta\theta^\mathrm{Q} \geq \Big[2\sqrt{\nu} N\Big]^{-1}.
\end{equation}
The right-hand side is now the Heisenberg limit, which can always be reached in the asymptotic limit of large $\nu$ \cite{GiovannettiReview}. 

In our photonic gear approach, Alice and Bob exchange photons in SAM-OAM superposition states [see Fig. \ref{QPlate} (c-d)]. Alice initially prepares $N$ horizontally-polarized photons, as in the classical strategy. However, before sending them to Bob, she first has them pass through a $q$-plate of charge $q$. The $q$-plate implements the bidirectional (unitary) mode transformations $\{a^{\dag}_{R,0} \leftrightarrow a^{\dag}_{L,-2q}, a^{\dag}_{L,0} \leftrightarrow a^{\dag}_{R,2q} \}$, where the subscripts $0$ and $\pm2q$ refer to the OAM values, and $a^{\dag}_{\pi,l}$ denotes the creation operator of a photon with polarization $\pi$ and OAM component $l$ \cite{Naga09}. This results in the following transformation of Alice's photons: $\ket{1}_{H,0} \stackrel{q-\text{plate}}{\longrightarrow}\frac{1}{\sqrt{2}}(\vert 1\rangle_{L,-2q}+\vert 1\rangle_{R,2q})$. Next, a half-wave plate (HWP) is used to invert the polarization, to obtain the transmitted states
\begin{equation}
\label{PsiG}
\ket{\Psi^\mathrm{C}_\mathrm{G}} = \frac{1}{\sqrt{2}}(\vert 1\rangle_{R,-2q}+\vert 1\rangle_{L,2q}).\\
\end{equation}
This single-photon state represents a superposition of $m=2q+1$ quanta in opposite total (spin $+$ orbital) angular momentum eigenmodes.  Likewise, if the HWP is removed, the case $m=2q-1$ is obtained.

%%%%% FIGURE 2 %%%%%
%%%%%%%%%%%%%%%%%%%%%%%%%%%%%%%%%%%%%%%%%%%%%%%%%%
\begin{figure*}[t]
\centering
\includegraphics[width=0.95\textwidth]{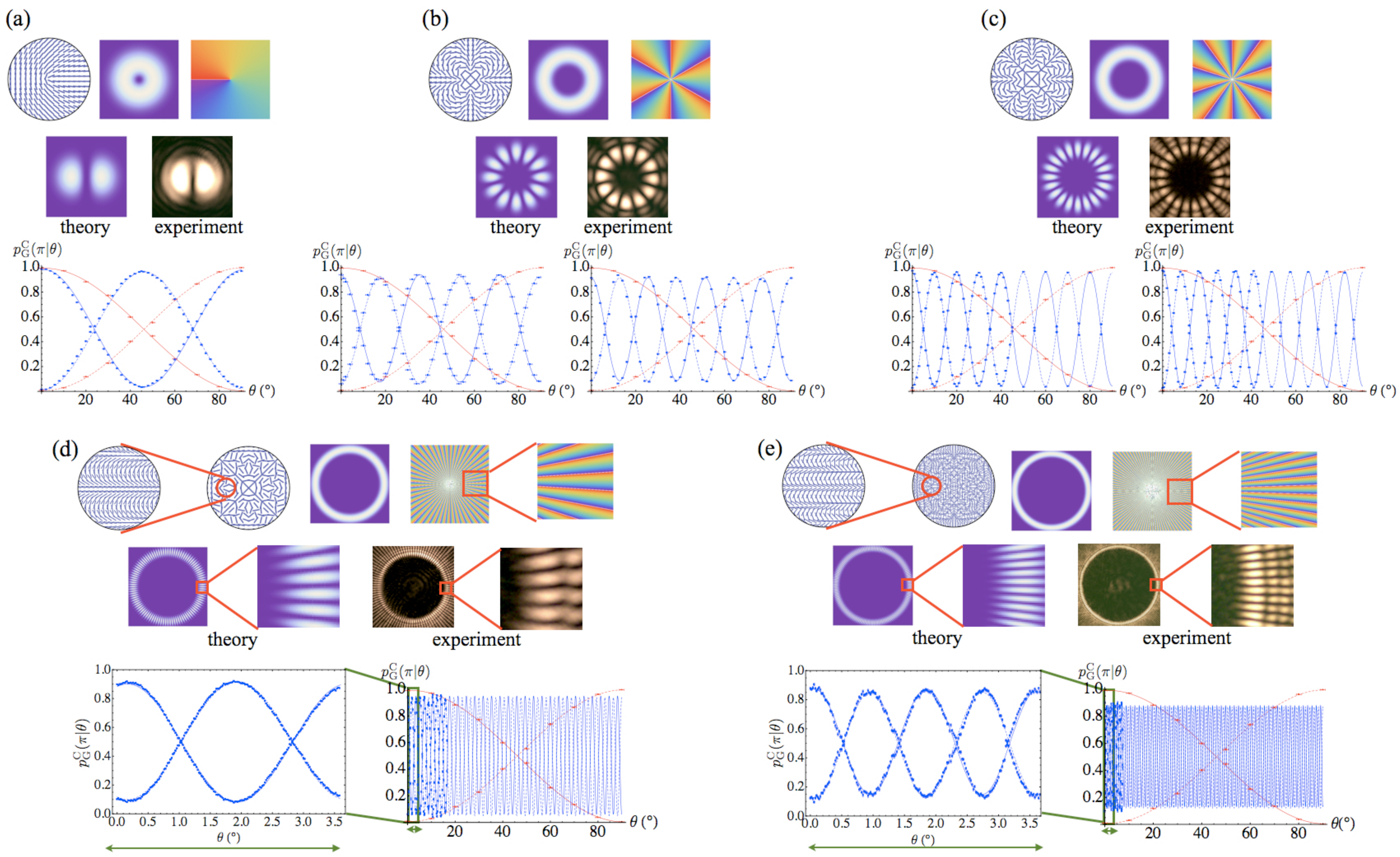}
\caption{\textbf{Single-photon experimental fringes}. Experimental results for single photons and $q$-plate charges (a) $q=1/2$ ($l=1$), (b) $q=3$ ($l=6$), (c) $q=5$ ($l=10$), (d) $q=25$ ($l=50$), and (e) $q=50$ ($l=100$). For each case (a-e), we report: the $q$-plate axis pattern, corresponding to the distribution of the liquid crystal molecular director (top row, left panels); the calculated intensity and phase profiles of the generated OAM fields (top row, middle and right panels); the theoretical and experimental intensities after projection on the $H$-polarization state (middle panels); the measured fringe patterns (blue dots) as a function of the mechanical rotation angle $\theta$, accompanied by sinusoidal best-fit curves (blue lines) and by the polarization-only case (red dots and lines) (bottom panels). The fringe patterns reported for cases (a,d-e) correspond to $m=2q+1$ [(a) $m=2$, (d) $m=51$, (e) $m=101$]. Fringe patterns with $m=2q-1$ (without the HWPs, left plots) and $m=2q+1$ (with the HWPs, right plots) are shown in (b) ($m=5$ and $m=7$) and (c) ($m=9$ and $m=11$). In (d) and (e), an inset with a zoomed-in region of the fringes is also shown. Error bars in the polarization fringes are due to the poissonian statistics of the recorded events, while error bars in the set value of the angle $\theta$ are due to the mechanical resolution of the rotation stage.}
\label{fig:experimental_fringes}
\end{figure*}
%%%%%%%%%%%%%%%%%%%%%%%%%%%%%%%%%%%%%%%%%%%%%%%%%%

In Bob's frame, the photons arrive as $\ket{\Psi^\mathrm{C}_\mathrm{G}(\theta)}=\frac{1}{\sqrt{2}}(e^{im\theta}\vert 1\rangle_{R,-2q}+e^{-im\theta}\vert 1\rangle_{L,2q})$. To detect them, he first undoes Alice's polarization flip with another HWP, and undoes her OAM encoding with another $q$-plate of the same charge $q$, so that
\begin{equation}
\label{eq:psi_G_final}
\ket{\Psi^\mathrm{C}_\mathrm{G}(\theta)}\longrightarrow \frac{1}{\sqrt{2}}(e^{i m\theta}\vert 1\rangle_{R,0} +e^{-i m\theta}\vert 1\rangle_{L,0}).\\
\end{equation}
This state corresponds to a uniform linear polarization, but with the polarization direction forming an angle $m\theta$ with respect to Bob's $H$ axis, resulting in the photonic gear effect. Finally, Bob measures the probability of detecting the $H$ linear polarization conditioned on $\theta$ as in the classical strategy. This is again given by Malus' law: 
\begin{equation}
\label{pg}
p^\mathrm{C}_\mathrm{G}(H|\theta)=\cos^2(m\theta), 
\end{equation}
but shows now the $m$-fold resolution enhancement over the polarization-only strategy.

As usual, Alice and Bob repeat the procedure a total of $\nu$ times. Their statistical error is now bounded as  
\begin{equation}
\label{DeltaCQ}
\Delta\theta^\mathrm{C}_\mathrm{G} \geq \Big[2m\sqrt{\nu N}\Big]^{-1},
\end{equation}
and can always saturate the bound in the asymptotic limit of large $\nu N$, as shown in Supplementary Note 1. This represents an improvement over the standard limit \eqref{DeltaC} for polarization-only strategies by a factor of $m$. This enhancement is not quantum but due exclusively to the coherent rotational sensitivity of high-order angular momentum eigenmodes.

%%%%% FIGURE 3 %%%%%
%%%%%%%%%%%%%%%%%%%%%%%%%%%%%%%%%%%%%%%%%%%%%%%%%%
\begin{figure*}[ht!]
\centering
\includegraphics[width=0.8\textwidth]{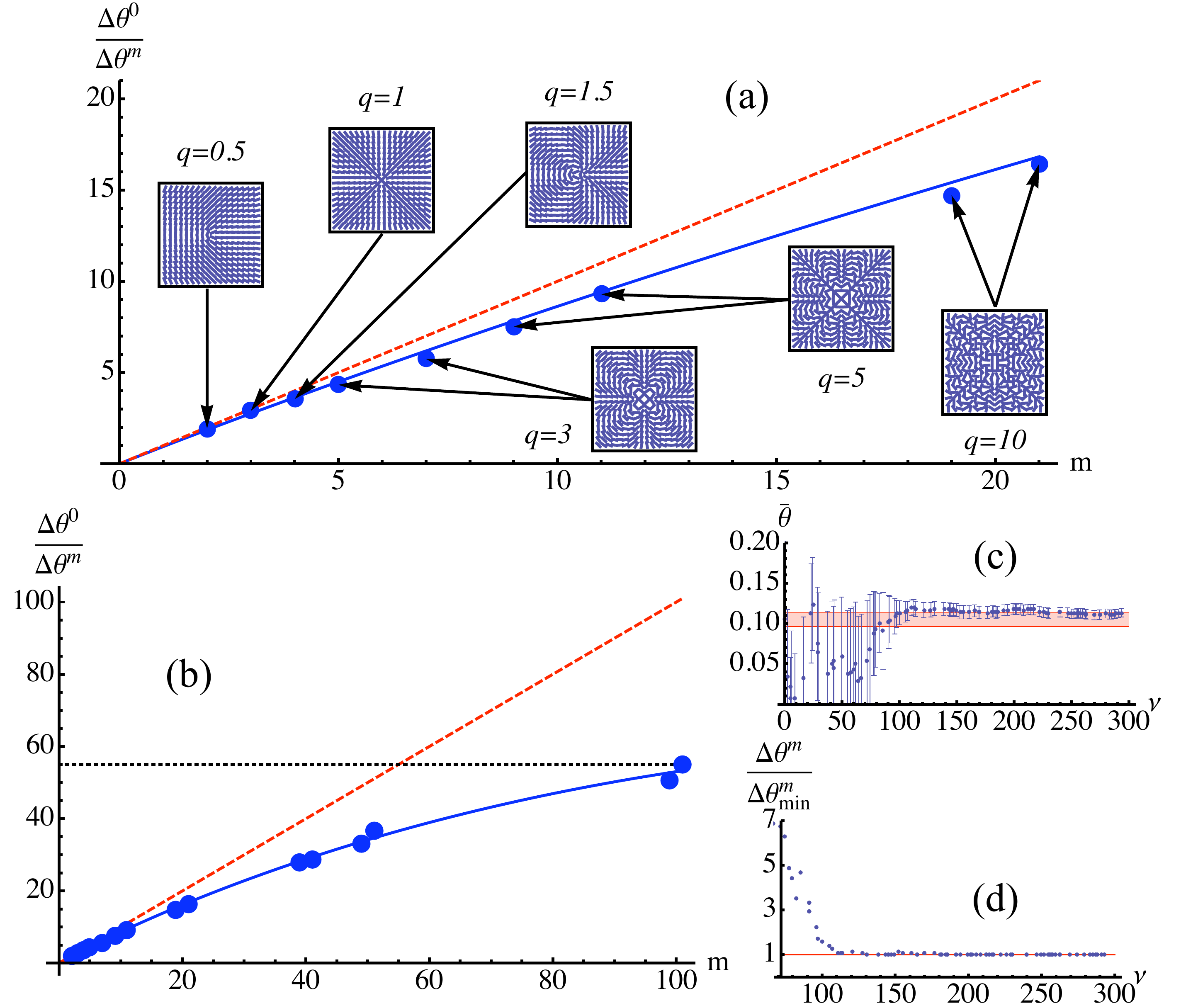}
\caption{\textbf{Estimation of a rotation angle with photonic gears in the single-photon regime.} Ratio between the statistical errors $\Delta \theta^{0}/\Delta \theta^{m}$ for the polarization-only strategy versus the gears strategy in the single-photon regime for $q<10$ (a) and for all the implemented values of $q$ (b). In (a) for each point the pattern of the corresponding $q$-plate is also shown. We obtain a maximum precision enhancement of $\approx 55$ for $q=50$, corresponding to the generation of optical states with an OAM component with $l=100$. Red dashed line: theoretical prediction for the ideal case. Blue solid line: model taking into account experimental imperfections [see Eq. (\ref{eq:model_main}) and Supplementary Note 1 for more details]. Black horizonal dashed line: experimental maximum enhancement. (c-d) Convergence of the angle estimation procedure as a function of the number of repeated experiments $\nu$ for $m=7$ ($q=3$). (c) Measured angle $\overline{\theta}$ versus the number of incident single photons $\nu$ (the red area corresponds to the true angle set in the apparatus, up to mechanical resolution). Error bars correspond to the statistical error in the estimation process. (d) Ratio $\Delta \theta^{m}/\Delta \theta_{\mathrm{min}}^{m}$ showing the convergence to the Cram\'er-Rao bound (see Supplementary Note 1).}
\label{fig:estimation}
\end{figure*}
%%%%%%%%%%%%%%%%%%%%%%%%%%%%%%%%%%%%%%%%%%%%%%%%%%

Remarkably, already for $m>\sqrt{N}$ the scaling \eqref{DeltaCQ} becomes better than the best precision \eqref{DeltaQ} attainable with polarization-only NOON states. Furthermore, the photonic-gear strategies, both in the single-photon and classical regimes, greatly outperform the latter in realistic scenarios with large $N$. First, the production and detection of our SAM-OAM photon states is exponentially more efficient in $N$ than those of NOON states. Second, since state \eqref{PsiG} does not bear any multi-photon coherences, losses reduce the total number of photons, but leave the remaining ones unaltered. That is, total losses characterized by an overall transmissivity $0\leq\eta\leq1$ enter as a constant multiplicative factor, simply rescaling in \eqref{DeltaCQ} the total number of photons to $\nu N \eta$ (see Supplementary Note 1), in striking contrast to NOON states \cite{Dorner}. Furthermore, one could also consider to exploit the orbital angular momentum of light to mimic the behaviour of other quantum states such us spin-squeezed state, requiring the superposition of many angular-momentum eigenstates.

{\bf Combining photonic gears and quantum enhancement}. The last type of strategy we consider is a hybrid classical-quantum one, which exploits both entanglement and high angular momenta through the photonic gear. In its simplest version, each probe may consist of an $N$-photon entangled NOON state $\ket{\Psi^\mathrm{Q}_\mathrm{G}}=\frac{1}{\sqrt{2}}(\vert N\rangle_{R,-2q}+\vert N\rangle_{L,2q})$. Following the same steps as above, one finds this time that  $p^\mathrm{Q}_\mathrm{G}(\Psi^{\mathrm{Q}}_\mathrm{G}|\theta)\doteq|\bra{\Psi^\mathrm{Q}_\mathrm{G}}\ket{\Psi^\mathrm{Q}_\mathrm{G}(\theta)}|^2=\cos^2(mN\theta)$ and  
\begin{equation}
\label{DeltaQG}
\Delta\theta^\mathrm{Q}_\mathrm{G} \geq \Big[2m\sqrt{\nu} N\Big]^{-1}.
\end{equation}
Thus, ideally, this strategy features the Heisenberg precision scaling for hybrid SAM-OAM approaches, but  it bears in practice the same loss-sensitivity problems as the polarization-only quantum strategy. However, for small $N$, these problems can still be efficiently dealt with and interesting applications can be achieved, as we demonstrate below.

Moreover, multi-photon quantum states other than NOON states can also be combined with the photonic gears, obtaining other interesting effects. For example, let us consider two-photon polarization-entangled states, where one photon is sent to Alice and the other to Bob. Alice and Bob make local $H/V$-polarization analysis in their own rotating stages, which can be set at arbitrary angles $\theta_{\mathrm{A}}$ and $\theta_{\mathrm{B}}$. When $\theta_{\mathrm{A}}=\theta_{\mathrm{B}}=\theta$ the system can model two photons travelling in the same mode, subject to the same rotation, and hence yield results analogous to the NOON-state case discussed above for $N=2$. 

\begin{figure*}[t]
\includegraphics[width=0.85\textwidth]{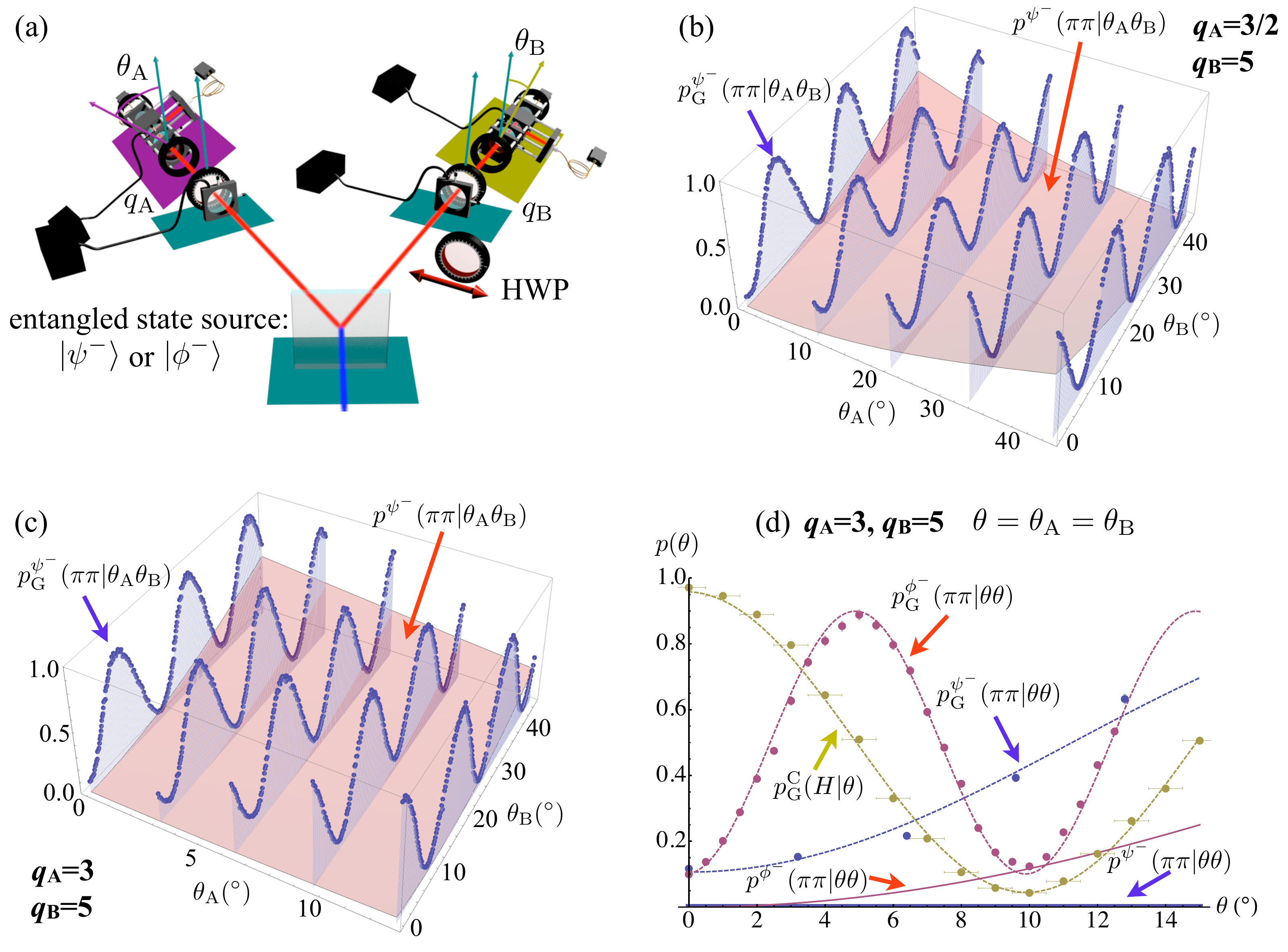}
\caption{\textbf{Entangled photonic gears.} (a) Experimental setup. An entangled photon pair in the polarization state $\vert \psi^{-} \rangle$ is generated by type-II spontaneous parametric down conversion. The state can be converted to the $\vert \phi^{-} \rangle$ state by inserting a HWP in the path of Bob's photon. Each photon is then converted into SAM-OAM hybrid states by the $q$-plates $q_{\mathrm{A}}$ and $q_{\mathrm{B}}$ and a HWP, as before, and is sent to a different rotation stage for the analysis. (b) Normalized experimental correlations $p^{\psi_-}_\mathrm{G}(\pi \pi|\theta_\mathrm{A},\theta_\mathrm{B})$ (blue points), with $\pi \pi = H H$, obtained with the $\vert \psi^{-} \rangle$ state by measuring the two-fold coincidences in the $H$-polarization bases on both modes for different values of the rotation angles $\theta_{\mathrm{A}}$ and $\theta_{\mathrm{B}}$. We observe the gear enhancement with respect to the polarization-only case (red surface, theory) in the oscillation frequencies in both directions $\theta_{\mathrm{A}}$ (with $m_{\mathrm{A}}=2 q_{\mathrm{A}}-1=2$) and $\theta_{\mathrm{B}}$ (with $m_{\mathrm{B}}=2 q_{\mathrm{B}}+1=11$). (c) Normalized experimental correlations again with $\vert \psi^{-} \rangle$ (blue points) but for $m_{\mathrm{A}}=2 q_{\mathrm{A}}+1=7$ and $m_{\mathrm{B}}=2 q_{\mathrm{B}}+1=11$. (d) Normalized experimental correlations obtained with the $\vert \psi^{-} \rangle$ and $\vert \phi^{-} \rangle$ states when rotating the two stages by the same angle $\theta_{\mathrm{A}}=\theta_{\mathrm{B}}=\theta$, for $m_{\mathrm{A}}=2 q_{\mathrm{A}}+1=7$ and $m_{\mathrm{B}}=2 q_{\mathrm{B}}+1=11$. The polarization correlations (blue points: data for $\vert \psi^{-} \rangle$, red points: data for $\vert \phi^{-} \rangle$) now present an oscillation pattern with a periodicity enhancement of $(m_{\mathrm{A}}+m_{\mathrm{B}})$ for $\vert \phi^{-} \rangle$ and $m_{\mathrm{A}}-m_{\mathrm{B}}$ for $\vert \psi^{-} \rangle$, due to quantum entanglement combined with the gear effect. The theoretical polarization-only $HH$ correlation (without the gear enhancement) are also shown, for reference, as a red solid curve in the $\vert \phi^{-}\rangle$ state case, oscillating as $2 \theta$, and as a blue solid curve in the $\vert \psi^{-}\rangle$ state case, which is constant and vanishing. Yellow points: experimental data for single-photon gear with $m=(m_{\mathrm{A}}+m_{\mathrm{B}})/2=9$, oscillating at half the frequency of $\vert \phi^{-}_{\mathrm{G}} \rangle$. Dashed curves: best fit of the experimental data. The visibility of the pattern for $\vert \phi^{-} \rangle$ state is $V^{\phi^{-}_{\mathrm{G}}}=0.826 \pm 0.011$. In all plots, error bars in the correlations are due to the poissonian statistics of the recorded events, while error bars in the set value of the angle $\theta$ are due to the mechanical resolution of the rotation stage.}
\label{fig:entanglement}
\end{figure*}

When $\theta_{\mathrm{A}}\neq\theta_{\mathrm{B}}$ one can instead align two distant frames remotely with two-photon probes produced by an unrelated common source, which sends one photon to each frame, by exploiting the quantum correlations among the two photons. More in detail, let us assume that the photons are generated in the maximally entangled polarization Bell state $\ket{\psi^{-}}=\frac{1}{\sqrt{2}}(\ket{1}_{R,0}^{\mathrm{A}}\ket{1}_{L,0}^{\mathrm{B}}-\ket{1}_{L,0}^{\mathrm{A}}\ket{1}_{R,0}^{\mathrm{B}})$. The photons, before transmission, are sent through two $q$-plates with topological charges $q_{\mathrm{A}}$ and $q_{\mathrm{B}}$, respectively, and a HWP, as shown in Fig.\ \ref{fig:entanglement}a. Thus, the following state is distributed to Alice and Bob:
\begin{equation}
\ket{\psi^{-}_\mathrm{G}}=\frac{1}{\sqrt{2}}(\ket{1}_{R,-2q_{\mathrm{A}}}^{\mathrm{A}}\ket{1}_{L,2q_{\mathrm{B}}}^{\mathrm{B}}-\ket{1}_{L,2q_{\mathrm{A}}}^{\mathrm{A}}\ket{1}_{R,-2q_{\mathrm{B}}}^{\mathrm{B}}).
\label{entangledgear}
\end{equation}
Alice and Bob, in their rotated frames, apply the same transformations to the photons, thus converting them back to pure polarization states. The probability that Alice and Bob both detect $H$-polarized photons in their local frames is then $p^{\psi_-}_\mathrm{G}(HH\vert \theta_{\mathrm{A}}\theta_{\mathrm{B}})=\frac{1}{2} \sin^{2}[(2q_{\mathrm{A}}+1)\theta_{\mathrm{A}}-(2q_{\mathrm{B}}+1)\theta_{\mathrm{B}}]$, showing ``amplified'' polarization correlations. Choosing $q_{\mathrm{A}}=q_{\mathrm{B}}$, one can for example use these correlations (in combination with classical communication channels) to precisely estimate the relative misalignment $\theta_{\mathrm{A}}-\theta_{\mathrm{B}}$ and remotely align the two distant frames. If a HWP with the optical axis parallel to $H$ is now inserted in Bob's photon path after generation of the polarization-entangled state (which corresponds to acting with a $\sigma_x$ Pauli operator in the $R/L$ basis), one obtains the entangled state $\vert \phi^{-}_\mathrm{G} \rangle =\frac{1}{\sqrt{2}}(\ket{1}_{L,2q_{\mathrm{A}}}^{\mathrm{A}}\ket{1}_{L,2q_{\mathrm{B}}}^{\mathrm{B}}-\ket{1}_{R,-2q_{\mathrm{A}}}^{\mathrm{A}}\ket{1}_{R,-2q_{\mathrm{B}}}^{\mathrm{B}})$, instead of $\ket{\psi^{-}_\mathrm{G}}$. Alice's and Bob's $HH$-photon correlations have now probability $p^{\phi^{-}}_\mathrm{G}(HH\vert \theta_{\mathrm{A}}\theta_{\mathrm{B}})=\frac{1}{2}\sin^{2}[(2q_{\mathrm{A}}+1)\theta_{\mathrm{A}}+(2q_{\mathrm{B}}+1)\theta_{\mathrm{B}}]$. So, in this case, for $\theta_{\mathrm{A}}=\theta_{\mathrm{B}}=\theta$, the system is metrologically equivalent to NOON state probes $\ket{\Psi^\mathrm{Q}_\mathrm{G}}$ for $N=2$ and $m=\frac{m_{\mathrm{A}}+m_{\mathrm{B}}}{2}=q_{\mathrm{A}}+q_{\mathrm{B}}+1$. In particular, for $N=2$ photons, $\theta$ can be estimated from the $HH$-correlation measurements with just half the efficiency as from $p^\mathrm{Q}_\mathrm{G}(\Psi^{\mathrm{Q}}_\mathrm{G}|\theta)$, which would require two-photon interference detection. Full efficiency in the estimation process can be recovered by simply registering and considering the four possible two-photon polarization-correlations ($HH, HV, VH, VV$), which requires no extra measurements. The same result can be generalized to $N$-photon entangled states.

%%%%% FIGURE 4 %%%%%
%%%%%%%%%%%%%%%%%%%%%%%%%%%%%%%%%%%%%%%%%%%%%%%%%%
%%%%%%%%%%%%%%%%%%%%%%%%%%%%%%%%%%%%%%%%%%%%%%%%%%

{\bf Experimental single-photon gear enhancement}. Our theoretical predictions were experimentally tested by exploiting a series of $q$-plates with increasing charge $q$. We focus first on the single-photon regime. The experimental setup is shown in Fig.\ \ref{QPlate} (e). Figure \ref{fig:experimental_fringes} shows the polarization fringes obtained for several values of $q$, corresponding to the ``super-resolving'' Malus' law \eqref{pg}. The red curves correspond to the polarization-only approach ($q=0$), shown for comparison.  The oscillation frequency $\propto m=(2q\pm1)$ highlights the improving angular resolution for increasing $q$. In Supplementary Figure S2, we also show that the initial phase of the oscillation can be tuned by choosing the appropriate input polarization state, and that this allows one in turn to optimize the sensitivity for any angle $\theta$.

Experimental imperfections lead to a non-unitary fringe visibility. As shown in Supplementary Note 1, the loss of visibility increases the statistical error as:
\begin{equation}
\label{eq:model_main}
\Delta \theta^{m} \geq \Big[2 m V_{m} \sqrt{\eta_{m}} \sqrt{\nu N}\Big]^{-1} = \Delta \theta^{m}_{\mathrm{min}},
\end{equation}
where $V_m$ is the visibility of the oscillation pattern and $\eta_m$ the efficiency of the detection system. In our case, all curves  show a visibility greater than $0.73$. As a figure of merit for the enhancement in precision, we consider the ratio between the statistical error of the polarization-only strategy and of the photonic gear: $\Delta \theta^{0}/\Delta \theta^{m}\propto m V_m \sqrt{\eta_m/\eta_0}$.  Figures \ref{fig:estimation} (a) and (b) show $\Delta \theta^{0}/\Delta \theta^{m}$ as a function of $m$ obtained from the interference curves. We obtain a maximum enhancement over the polarization-only strategy by $\approx$ 55.  To obtain the same precision $\Delta \theta$ with the polarization-only classical strategy, one would have to increase the number of trials by a factor of $55^2=3025$, while for the polarization-only quantum NOON-state strategy, one would require entangled states of $N\approx 55$ photons each.  As shown in Figure \ref{fig:estimation} (c)-(d), our estimation protocol gives an estimate $\overline{\theta}$ which converges to the true value $\theta$ in a limited number of trials $\nu \sim 300$, where $\nu$ is the number of single photons sent through the system. Furthermore, in Supplementary Figure S2 and Supplementary Note 2, we discuss a three-step adaptive protocol which permits efficient and unambiguous estimation of such $\theta$ even when it is a completely unknown rotation in the full $[0, 2\pi)$ interval.

In Supplementary Figure S3, we show that the rotational sensitivity enhancement due to the photonic gears effect can also be achieved in the classical regime with an intense laser, making it immediately applicable to real-world optical measurements, which we will now briefly discuss. There, the most common problem is to perform precise non-contact and/or remote optical measurements of roll angles.  These are mechanical rotations of an object around one of its symmetry axes \cite{Liu03,Li05}. Polarization-based methods, essentially relying on the Malus' law combined with suitable polarization manipulations, are among the most convenient approaches. Depending on the details of the scheme, this typically leads to a sensitivity of about ${10^{-2}}$ degrees for a dynamical range of  $30-360^\circ$, or about ${10^{-4}}$ degrees when restricting the range to $\sim1^{\circ}$. All these polarization-based methods, irrespective of the details, can be combined with our photonic gear tool without changes. Their sensitivity is therefore predicted to be improved approximately by the factor $m\times V_m$, which we have shown can be made larger than 50. For example, the method reported in Ref.\ \onlinecite{Liu03} combined with our photonic-gear enhancement is expected to achieve a maximal sensitivity of ${10^{-6}}$ degrees, or about 0.01 arcsec. The dynamical range is also reduced by a similar factor, but the full dynamical range can be recovered by the adaptive protocol discussed in Supplementary Note 2.

{\bf Experimental photonic gears on two-photon entangled states}. We consider at last the quantum regime of entangled photons, using the setup shown in Fig. \ref{fig:entanglement} (a). We demonstrate two-photon entangled states where each photon has a different total angular momentum, $m_1$ and $m_2$, with a maximum of $m_1+m_2=18$. We carried out two types of experiments. In the first, we generated photon pairs in the ``entangled photonic gear state'' $\ket{\psi^{-}_\mathrm{G}}$, given in Eq.\ (\ref{entangledgear}). We then measured the $HH$ correlations for two different sets of $q$-plates. The results are reported in Fig. \ref{fig:entanglement} (b-d) as a function of the angles $\theta_{\mathrm{A}}$ and $\theta_{\mathrm{B}}$ of Alice's and Bob's stages. The enhancement in oscillation frequency in both the $\theta_{\mathrm{A}}$ and the $\theta_{\mathrm{B}}$ directions with respect to the polarization-only case is clearly observed and matches our theoretical predictions. Next, we generated the entangled state $\vert \phi^{-}_\mathrm{G} \rangle$ and rotated the two stages by the same angle $\theta_{\mathrm{A}}=\theta_{\mathrm{B}}=\theta$, thus creating a situation analogous to the case of NOON state probes. The measurement results are shown in Fig.\ \ref{fig:entanglement} (d). The hybrid quantum-classical sensitivity enhancement by the factor $m N=m_{\mathrm{A}}+m_{\mathrm{B}}=2q_{\mathrm{A}}+2q_{\mathrm{B}}+2$ is clearly observed, confirming again our predictions. In particular, the experimental comparison between the 2-photon quantum case with $\vert \phi^{-}_{\mathrm{G}}\rangle$ and the single-photon case with $\vert \Psi^{\mathrm{C}}_{\mathrm{G}} \rangle$ shows that in the former case a quantum enhancement by a factor 2 is superimposed to the classical photonic gear effect. 

\section*{Discussion}

In summary, we have reported a  photonic scheme to measure rotation angles with greatly enhanced precision. In the  regime of single-photon probes, a precision of $\sim 55 \sqrt{\nu N}$ has been demonstrated experimentally, with $\nu N$ the total number of photons. Notably, rather than in an asymptotic limit, this precision was attained already for total photon numbers as small as $\nu N \approx 10^2$ to $10^4$. To our knowledge, this constitutes the highest precision per-particle reported so far \cite{Leibfried05, Nagata07, Jones09, Higgins07}. In addition, we demonstrated ultra sensitive two-photon entangled probes tailored for different target estimations. We anticipate that immediate application of the photonic-gear concept in a classical regime can improve current polarization-based methods for measuring roll angles to a sensitivity of less than 0.01 arcsec. These values provide substantial progress over the current state of the art. In addition to metrological applications, the capability we have demonstrated to efficiently generate and detect hybrid polarization-OAM quantum states with very large OAM creates interesting prospects for high-rate classical communication \cite{Wang12}, coupling with atomic systems \cite{Ande06}, quantum information processing \cite{Dada11}, and fundamental tests of quantum mechanics \cite{Naga12,DAmb13}.

\section*{Methods}

{\bf Q-plates fabrication}. A $q$-plate is essentially a half-wave plate whose optical axis orientation angle is not uniform, but changes from point to point in the transverse plane. The optical axis orientation at each point can be given as $\theta=q\varphi+\alpha$, where theta is the angle the optical axis forms with a reference axis $x$ in the transverse plane $xy$, $\varphi$ is the azimuthal angle coordinate, $q$ is called topological charge and $\alpha$ is a constant. When a circularly polarized photon passes through a $q$-plate, its helicity is switched to the opposite one, like in the case of standard half-wave plates. Such polarization transformation, while having the same initial and final states, occurs in a different way in different points of the transverse plane, giving rise to a non-uniform geometrical (or Pancharantam-Berry) phase retardation. The phase shift is coordinate dependent and equal to the double of theta, which is equivalent to an OAM state change of $\pm2q$. The sign of the phase shift depends on the initial polarization state: for the circular $\ket{1}_{L,m}$ input, the output OAM state is changed by $+2q$ and for the circular $\ket{1}_{R,m}$ one, the output OAM state is changed by $-2q$. The same applies to all the superposition states that correspond to an elliptically polarized input photons, producing hybrid SAM-OAM states in the output. This way, with a single $q$-plate of charge $q$ and linearly polarized Gaussian input it is possible to generate a hybrid state $\frac{1}{\sqrt{2}}(\vert 1\rangle_{L,-2q}+\vert 1\rangle_{R,2q})$.

In our experiment, the $q$-plates are realized using a nematic liquid crystal (LC). A $q$-plate is a LC cell, in which a thin layer of LC is sandwiched between two glass windows, previously covered with a suitable surface coating for inducing the desired orientational order to the LC layer. For generating the $q$-plate pattern, we adopted a photoalignment method \cite{chigrinov_book}, which consists in exposing the aligning layer, which contains suitable photosensitive dyes, with linearly polarized UV light. The polarization direction of the light defines the local anisotropy direction of the aligning coating (the latter is actually orthogonal or parallel to the polarization, depending on the choice of materials), which in turn, induces the orientation direction of the LC layer. Hence, the $q$-plate is realized by illuminating the empty cell with a thin line of light, ``writing'' sector-by-sector the desired angular structure, while constantly controlling the polarization direction of the writing beam and rotating the sample at the same time. The LC is inserted in the cell after writing the alignment pattern. The cell glasses include also a thin conducting transparent layer of Indium Tin Oxide (ITO), in order to apply an electric field and fine-tune the $q$-plate to the desired total retardation of half-wave.
The output OAM state quality is defined by the smoothness of the $q$-plate and the size of central defect. The first one is defined by the sample rotation step, laser line thickness and the choice of the aligning surfactant material. The central defect size is also very sensitive to the alignment of the sample  with respect to the writing beam. For low topological charges $q < 10$, a step of $1^\circ$ and corresponding laser line thickness provides good quality of the final structure. With increase of topological charge, finer $q$-plate structures required smaller rotation steps (up to $0.1^\circ$ for $q\geq25$) and particular attention to rotation error compensation. In combination with suitable azodyes selected as high resolution aligning surfactants \cite{chigrinov_book}, it was possible to achieve a high quality of the LC alignment and a low size, with respect to the total aligned area, of the central defect.

{\bf Experimental details}. The optical source of single photons (not shown) is provided by a Ti:Sa mode-locked laser, which generates pulses of $\Delta t = 120$ fs with repetition rate of $76$ MHz at $\lambda = 795$ nm. The source is then doubled in frequency via a second harmonic generation process to obtain the pump beam of the experiment of $\lambda = 397.5$ nm and $P = 700$ mW. The single-photon is conditionally generated through a type-II spontaneous parametric down conversion process in a $l=1.5$ mm beta-barium-borate (BBO) crystal. The output field is then filtered in frequency with a $\Delta \lambda = 3$ mm filter (IF) and coupled in a single-mode fiber.  Photons at the output of this fiber are then injected into Alice's preparation device. 

%%%%%%%%%%%%%%%%%%%%%%%%%%%%%%%%%%%%%%%%%%%%%%%%%%

\section*{Acknowledgements}

This work was supported by the FET-Open Program, within the 7th Framework Programme of the European Commission under Grant No. 255914, PHORBITECH. V.D., N.S., L.D., F.S. acknowledge support from FIRB-Futuro in Ricerca (HYTEQ), Progetto Ateneo of Sapienza Universit\`a di Roma, Finanziamento per Avvio alla Ricerca 2012 of Sapienza Universit\`a di Roma. S.P.W. acwnowledge support from the brazilian agencies CNPq and FAPERJ. L.A. acknowledge support from the Spanish Juan de la Cierva foundation and the EU under Marie Curie IEF No 299141. Y.L. and L.C.K. acknowledge support from the National Research Foundation and the Ministry of Education in Singapore.  

\section*{Author contributions}

L.A., Y.L. and L.C.K. had the initial idea of exploiting NOON-like photonic states of OAM for ultrasensitive angular metrology. All the authors contributed to developing the concept, the theory, and the classical and quantum applications. S.S. and L.M. developed the high-charge q-plates. V.D., N.S., L.D. and F.S. carried out the experiments. V.D., N.S., L.D., F.S., S.P.W., and L.A. analyzed the data. All the authors contributed to writing the manuscript.

\section*{Additional information}

{\bf Competing financial interests} All authors are listed as inventors on an Italian patent pending application (RM2013A000318, deposited 3 June 2013) entitled ``Ultra-sensitive photonic tiltmeter utilizing the orbital angular momentum of the light, and relevant angular measurement method'' dealing with the application of the photonic gears to angular measurements.

\end{document}

\title{Photonic polarization gears for ultra-sensitive angular measurements - \\
Supplementary Information}

\author{Vincenzo D'Ambrosio}
\affiliation{Dipartimento di Fisica, Sapienza Universit\`a di Roma, piazzale Aldo Moro 5, 00185 Roma, Italy}
\author{Nicol\`o Spagnolo}
\affiliation{Dipartimento di Fisica, Sapienza Universit\`a di Roma, piazzale Aldo Moro 5, 00185 Roma, Italy}
\author{Lorenzo Del Re}
\affiliation{Dipartimento di Fisica, Sapienza Universit\`a di Roma, piazzale Aldo Moro 5, 00185 Roma, Italy}
\author{Sergei Slussarenko}
\affiliation{Dipartimento di Fisica, Universit\`a di Napoli Federico II, Compl. Univ. di Monte S. Angelo, Napoli, Italy}
\author{Ying Li}
\affiliation{Centre for Quantum Technologies, National University of Singapore, Singapore 117543}
\author{Leong Chuan Kwek}
\affiliation{Centre for Quantum Technologies, National University of Singapore, Singapore 117543}
\affiliation{Institute of Advanced Studies (IAS), Nanyang Technological University, Singapore 639673}
\affiliation{National Institute of Education, Nanyang Technological University, Singapore 637616}
\author{Lorenzo Marrucci}
\affiliation{Dipartimento di Fisica, Universit\`a di Napoli Federico II, Compl. Univ. di Monte S. Angelo, Napoli, Italy}
\affiliation{CNR-SPIN, Complesso Universitario di Monte S. Angelo, 80126 Napoli, Italy}
\author{Stephen P. Walborn}
\affiliation{Instituto de F\'{i}sica, Universidade Federal do Rio de Janeiro, Rio de Janeiro, RJ 21941-972, Brazil}
\author{Leandro Aolita}
\affiliation{ICFO-Institut de Ci\`encies Fot\`oniques, Parc Mediterrani
 de la Tecnologia, 08860 Castelldefels (Barcelona), Spain}
\affiliation{Dahlem Center for Complex Quantum Systems, Freie Universit\"{a}t Berlin, Berlin, Germany}
\author{Fabio Sciarrino}
\email{fabio.sciarrino@uniroma1.it}
\affiliation{Dipartimento di Fisica, Sapienza Universit\`a di Roma, piazzale Aldo Moro 5, 00185 Roma, Italy}
\affiliation{Istituto Nazionale di Ottica (INO-CNR), Largo E. Fermi 6, I-50125 Firenze, Italy}

%%%%%%%%%%%%%%%%%%%%%%%%%%%%%%%%%%%%%%%%%%%%%%%%%%
\maketitle
%%%%%%%%%%%%%%%%%%%%%%%%%%%%%%%%%%%%%%%%%%%%%%%%%%

%
\begin{figure}[ht!]
\centering
\includegraphics[width=0.45\textwidth]{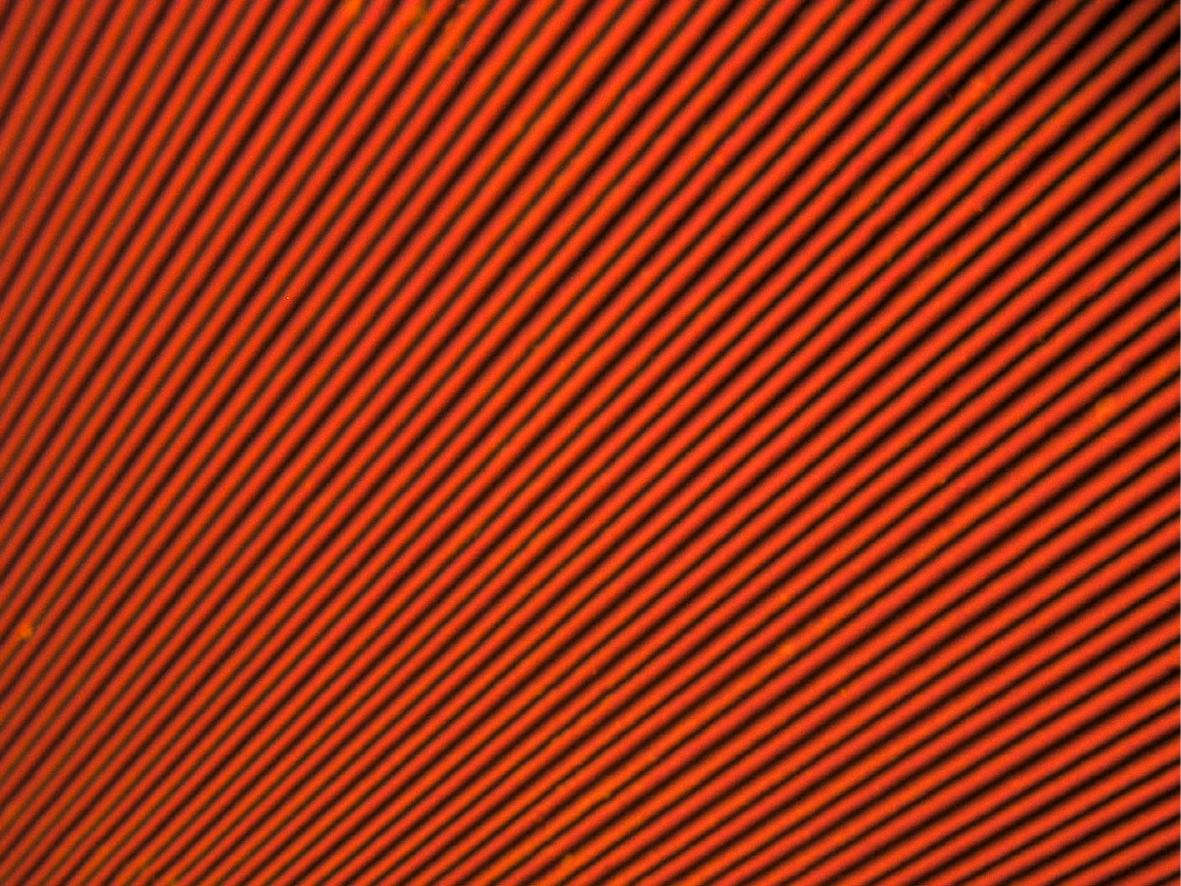}
\caption{}
\label{fig:qplate_photo}
\end{figure}
\vspace{-1.cm}
\noindent {\bf Supplementary Figure S1: Polarization microscope image of a $q$-plate.} The image shows an area of $q$-plate with $q=150$, located approximately $1$ mm from the center. Image size is $366$ $\mu$m in horizontal and $275$ $\mu$m in vertical.

\newpage

\begin{figure}[ht!]
\centering
\includegraphics[width=0.85\textwidth]{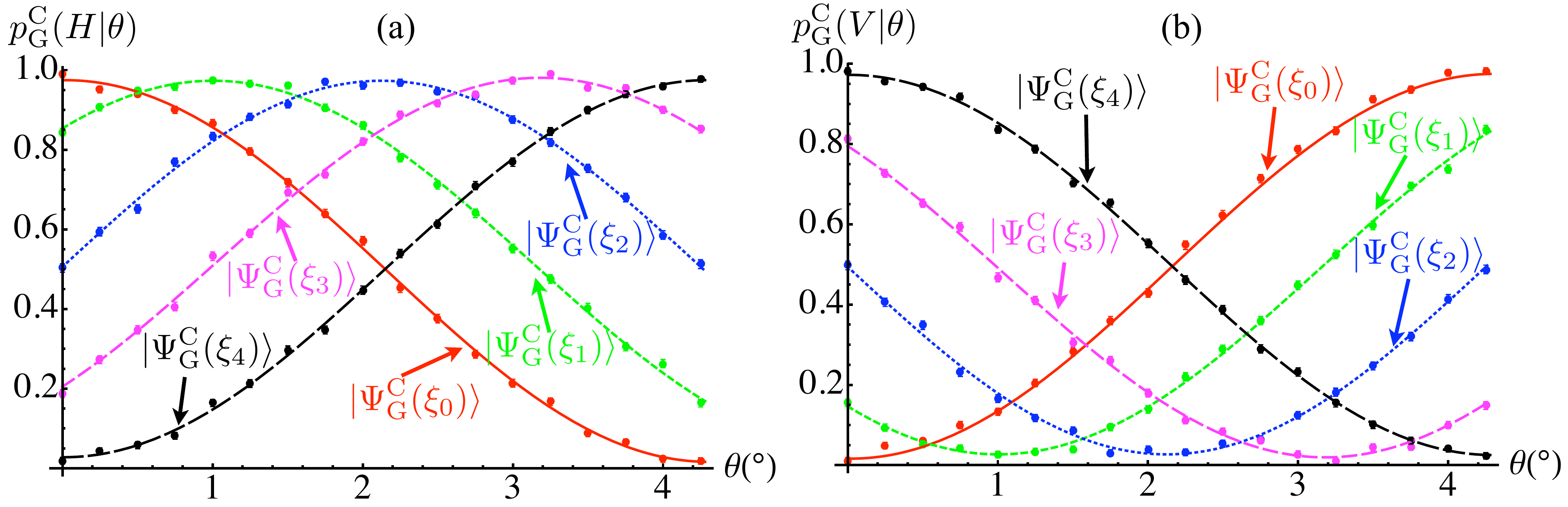}
\caption{}
\label{fig:fringe_adaptive}
\end{figure}
\vspace{-1.cm}
\noindent {\bf Supplementary Figure S2: Experimentally adapting the oscillation phase}. In order to implement the three-step protocol described in Supplementary Note 2 for the estimation of a completely unknown rotation angle, it is necessary to adapt the relative phase $\xi$ in the state $\vert \Psi_{\mathrm{G}}^{\mathrm{C}} (\xi) \rangle$ at each step of the process. This can be efficiently implemented by rotating the polarization of the input state of an angle $\xi$ by means of a half waveplate. Indeed, a rotation of the input state from $\vert 1 \rangle_{H}$ to $\vert 1 \rangle_{V}$ corresponds to an inversion in the maxima and the minima of the pattern. Here we report the oscillation patterns [(a) $p_{\mathrm{G}}^{\mathrm{C}}(H \vert \theta)$ and (b) $p_{\mathrm{G}}^{\mathrm{C}}(V \vert \theta)$] obtained with the single-photon photonic gear ($q=10$ and $m=2q+1=21$) for different values of initial state phase $\xi_{i}$: $\{\xi_{0}=0,\xi_1=\pi/8,\xi_2=\pi/4,\xi_3=3\pi/8,\xi_4=\pi/2\}$. Points: experimental data. Curves: corresponding best-fit curves. Error bars in the experimental data are due to the Poissonian statistics of the measured single-photon counts.
\newpage

\begin{figure}[ht!]
\centering
\includegraphics[width=0.4\textwidth]{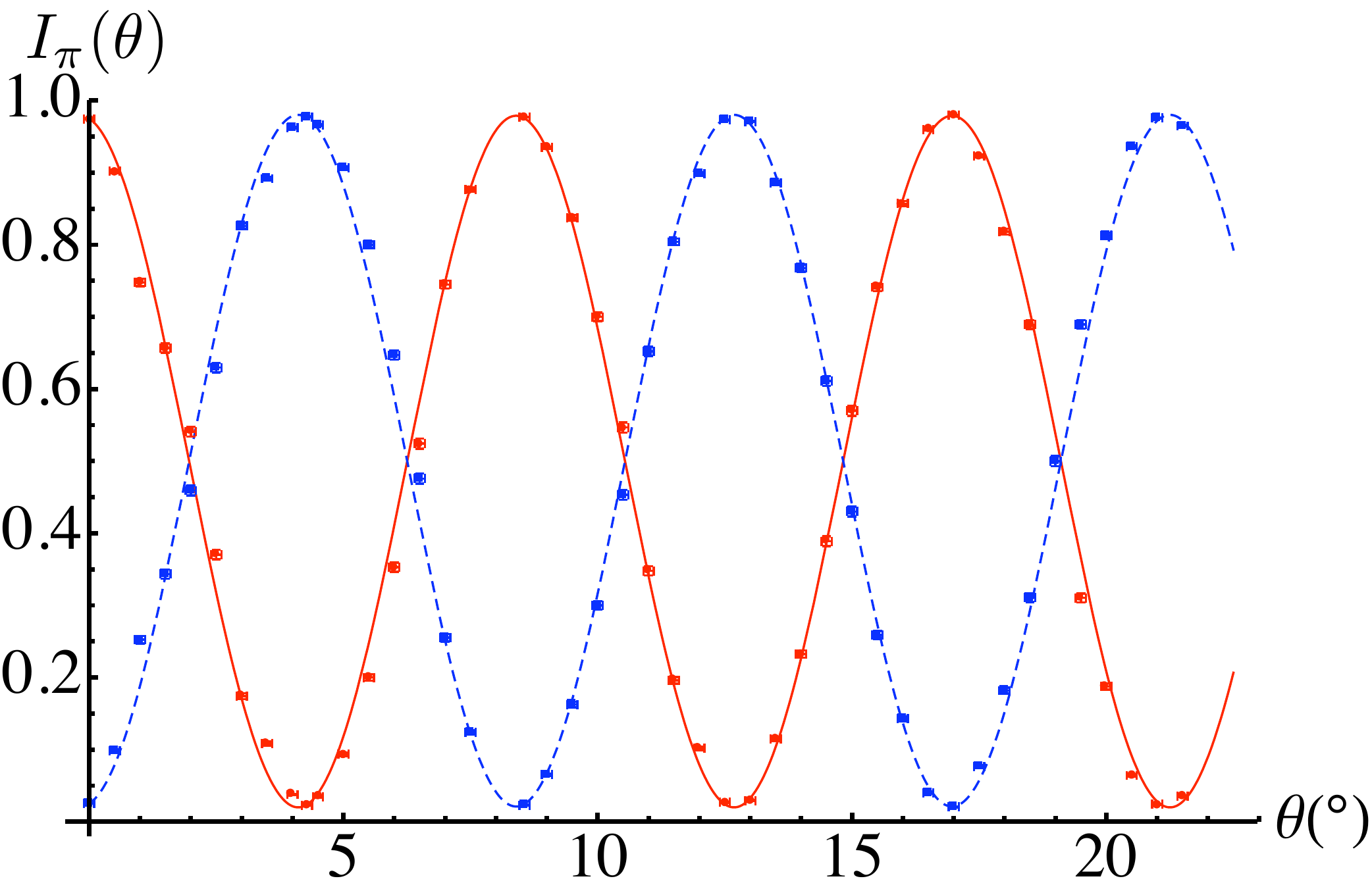}
\caption{}
\label{fig:fringe_coherent}
\end{figure}
\vspace{-1.cm}
\noindent {\bf Supplementary Figure S3: Classical regime hybrid photonic gears}. The photonic gear concept based on hybrid polarization-OAM states can be adopted also with intense light in the classical regime. To this end, we prepared as input a $P=1$ mW laser pulse with linear $H$ polarization.  The output state is analyzed by measuring the intensity in the two orthogonal linear $H$ and $V$ polarization directions. Here we report the measured oscillation patterns for classical intense input light with a value $q=10$ and $m=2q+1=21$. We observe that the super-resolution effect due to the hybrid encoding is efficiently observed also when the input state is a coherent classical beam. Indeed, the experimental visibilities for the two output intensities $I_{H}(\theta)$ (red points and curves) and $I_{V}(\theta)$ (blue points and curves) in this case are respectively $V_{H} = 0.95 \pm 0.02$ and $V_{V} = 0.91 \pm 0.02$. Points: experimental normalized intensities. Curves: corresponding best-fit curves. Error bars are due to fluctuations in the measured laser intensity.

\newpage

%%%%%%%%%%%%%%%%%%%%%%%%%%%%%%%%%%%%%%%%%%%%%%%%%%%%%%%%%%%%%%%%%%
\section*{Supplementary Note 1: Photonic gears and their precision}
\label{SIsinglephoton}

In this note, we describe the theoretical details behind the bound given in Eq. (6) of the main article for the precision of the hybrid SAM-OAM photonic gear concept.

\subsection{Quantum Fisher information including photon-losses and non-unitary conversion efficiency} 
\label{QFisherAppend}
The maximum amount of information which can be extracted about a parameter $\theta$ from an $N$-photon state $\varrho_N(\theta)$ is quantified by the quantum Fisher information $F_{N}=F_{N}\big(\varrho_N(\theta)\big)$. In particular, the minimal uncertainty in the estimation  of $\theta$ from measurements on $\varrho_N(\theta)$, optimized over all possible measurements, satisfies the so-called quantum Cram\'{e}r-Rao (QCR) bound: 
\begin{equation}
\Delta \theta_{\text{min}} \geq \big(\nu F_{N}\big)^{-1/2},
\label{eq:QCRbound}
\end{equation}
where $\nu$ is the number of measurement runs with copies of $\varrho_N(\theta)$. In addition, the bound can always be saturated in the asymptotic limit of large $\nu$, for a suitable estimation function, as discussed in Supplementary Note 2. See Ref. 1 for a review. 

When $\varrho_N(\theta)$ is obtained from a unitary evolution $\mathcal{U}_{\theta} = e^{-\imath \mathcal{G} \theta}$ of a pure state $\vert \psi_N(\theta) \rangle = \mathcal{U}_{\theta} \vert \psi^0_N \rangle$, the quantum Fisher information takes the simple $\theta$-independent form: $F_{N}\big(\varrho_N(\theta)\big)=F_{N}(\psi^0_N)= 4 \langle \psi^0_N \vert (\Delta \mathcal{G})^2 \vert \psi^0_N \rangle$, where $\Delta \mathcal{G}=\mathcal{G}-\langle \psi^0_N \vert\mathcal{G}\vert \psi^0_N \rangle$. For the hybrid SAM-OAM strategy with single photons, each photon is prepared in the state $\vert \Psi_{\mathrm{G}}^{\mathrm{C}} \rangle$. Then, in the absence of losses,  the state $\vert \psi^0_N \rangle$ is given by the $N$-photon state $\ket{\Psi_{\mathrm{G}}^{\mathrm{C}}}^{\otimes N}$.  The generator of the rotations around the $z$ axis is  $\mathcal{G}=(S^z_{\text{p}} + S^z_{\text{o}})/\hbar$, where $S^z_{\text{p}}$ and $S^z_{\text{o}}$  are are the corresponding  polarization and orbital angular momentum operators, respectively. The quantum Fisher information is additive, hence $F_{N}(\ket{\Psi_{\mathrm{G}}^{\mathrm{C}}}^{\otimes N})= N F$, where we have introduced the notation $F=F_1(\ket{\Psi_{\mathrm{G}}^{\mathrm{C}}})$. By evaluating the variance of $\mathcal{G}$ over the unperturbed single-photon state $\vert \Psi_{\mathrm{G}}^{\mathrm{C}} \rangle$, we obtain the explicit expression $F=4m^2$ for the  Fisher information per photon. This, together with the QCR bound above, gives us  precisely bound (6) of the main text:
\begin{equation}
\label{eq:QCR}
\Delta \theta_{\mathrm{G}}^{\mathrm{C}}\geq \frac{1}{2 m \sqrt{\nu N}}.
\end{equation}

Let us now consider the presence of photon losses. The total losses (including those at the detection stage) are quantified by a single transmittivity parameter $0\leq\eta\leq 1$, and transform the ideal single-photon state $\ket{\Psi_{\mathrm{G}}^{\mathrm{C}}(\theta)}$ arriving at Bob's station into:
\begin{equation}
\label{eq:state_losses}
\varrho_{\mathrm{G}}^{\mathrm{C}}(\theta) = \eta \vert \Psi_{\mathrm{G}}^{\mathrm{C}}(\theta) \rangle \langle \Psi_{\mathrm{G}}^{\mathrm{C}}(\theta) \vert + (1-\eta) \vert 0 \rangle \langle 0 \vert,
\end{equation}
where $\vert 0 \rangle$ is the vacuum state. Since $\varrho_{\mathrm{G}}^{\mathrm{C}}(\theta)$ is the sum of two terms which correspond to orthogonal subspaces, the quantum Fisher information per photon is
\begin{equation}
F\big(\varrho_{\mathrm{G}}^{\mathrm{C}}(\theta)\big)= \eta F(\vert\Psi_{\mathrm{G}}^{\mathrm{C}}(\theta) \rangle)+ (1-\eta) F(\vert 0 \rangle) =\eta F.
\end{equation}
Here we have used the fact that $F(\vert 0 \rangle)=0$, since no information can be acquired from the vacuum state. Using the additivity property of the Fisher information, the QCR bound \eqref{eq:QCRbound}, the precision in presence of losses is
\begin{equation}
\label{eq:QCR_losses}
\Delta \theta_{\mathrm{G}}^{\mathrm{C}}\geq \frac{1}{2 m \sqrt{\eta \nu N}}.
\end{equation}
That is, the overall effect of photon losses is simply to rescale the total number of photons from $\nu N$ to $ \eta \nu N$. 

We conclude by discussing the effect of a non-unitary conversion efficiency of the $q$-plates. In this case, the action of these devices are described by the following map:
\begin{equation}
\mathcal{Q}(\varrho) = \epsilon \, \mathcal{U}_{q} \varrho \mathcal{U}^{\dag}_{q} + (1-\epsilon) \mathbbm{1}_{\pi,q} \varrho \mathbbm{1}_{\pi,q},
\end{equation}
where $\mathcal{U}_{q}$ is the unitary evolution operator representing a perfect $q$-plate, and $\mathbbm{1}_{\pi,q}$ is the identity operator on the joint polarization-OAM space, and $\epsilon$ is the conversion efficiency. The output state of a photon after the encoding, rotation and decoding stages is measured with multimode fibers. The multimode fiber accepts all OAM states, and thus no measurement of the final OAM is performed, corresponding mathematically to the partial trace on the OAM subspace. The output density matrix then reads:
\begin{equation}
\varrho_{\mathrm{G}}^{\mathrm{C}}(\theta) = \epsilon_{1} \epsilon_{2} \vert \Psi_{\mathrm{G}}^{\mathrm{C}}(\theta) \rangle \langle \Psi_{\mathrm{G}}^{\mathrm{C}}(\theta) \vert +[\epsilon_{1}(1-\epsilon_{2}) + (1-\epsilon_{1})\epsilon_{2}] \frac{\mathbbm{1}_{\pi}}{2} + (1-\epsilon_{1})(1-\epsilon_{2}) \vert \Psi^{\mathrm{C}}(-\theta) \rangle \langle \Psi^{\mathrm{C}}(-\theta) \vert,
\end{equation}
where $\mathbbm{1}_{\pi}=\vert 1 \rangle_{R} \, _{R}\langle 1 \vert + \vert 1 \rangle_{L} \, _{L}\langle 1 \vert $, and $\epsilon_1$ and $\epsilon_2$ refer respectively to the conversion efficiencies of Alice's and Bob's $q$-plates. Finally, we observe that the latter term in $(1-\epsilon_{1})(1-\epsilon_{2})$ is small for typical efficiencies of the $q$-plates $(\epsilon_{i}>0.9)$, and can be neglected. The output state before detection can be then written as:
\begin{equation}
\label{eq:state_complete}
\varrho_{\mathrm{G}}^{\mathrm{C}} (\theta) \simeq V \vert \Psi_{\mathrm{G}}^{\mathrm{C}}(\theta) \rangle \langle \Psi_{\mathrm{G}}^{\mathrm{C}}(\theta) \vert + (1-V) \frac{\mathbbm{1}_{\pi}}{2},
\end{equation}
where $V = \epsilon_{1} \epsilon_{2}/[(1-\epsilon_{1}) (1-\epsilon_{2})]$ is the fringe pattern visibility achieved with polarization detection. Evaluating the quantum Fisher information for state (\ref{eq:state_complete}) again in the presence of photon losses, gives the bound for the precision $\Delta \theta^{m}$:
\begin{equation}
\label{eq:QCR_complete}
\Delta \theta^{m}\geq \frac{1}{2 m V \sqrt{\eta \nu N}}.
\end{equation}

In principle, $\eta$ and $\epsilon$ display no dependence on $q$ and therefore enter in the precision as just constant factors. In practice, a number of experimental imperfections can cause the two parameters to slightly depend on $q$. We considered this effect with an heuristic model, which includes a dependence on $q$ in Eq. (\ref{eq:QCR_complete}) of the form: $V \sqrt{\eta} \rightarrow V_{0} \sqrt{\eta}_{0} [1- \gamma m^{\delta}]$. The parameters $\gamma$ and $\delta$ are retrieved by fitting the experimental data, while $V_{0}$ and $\eta_{0}$ are the visibility and the efficiency for $q=0$. For the data shown in Figure 3 of the main text, we obtained $\gamma = 0.026 \pm 0.008$ and $\delta=0.62 \pm 0.07$, showing that the dependence on $q$ is very weak. 
%%%%%%%%%%%%%%%%%%%%%%%%%%%%%%%%%%%%%%%%%%%%%%%%%%%%%%%%%%%%%%%%%%%%%%%%%%%%%%%%%%%%%%%%%%%%%%%%%%%%%%%%%%%%%%%%%%%%%%%%%%%%%%%%%%%%%%%%%%%%%%%%%%%%

\subsection{Classical Fisher information of the outcomes of polarization measurements} 
Let us next show that the measurement strategy described in the main text is indeed the optimal one, reaching the QCR \eqref{eq:QCRbound}. We do this first for the ideal case $\eta=1$ and $V=1$, and then, below, briefly present the case with experimental imperfections. In the $H/V$ linear polarization basis, the state of each photon arriving at Bob's station reads
\begin{equation}
\ket{\Psi_{\mathrm{G}}^{\mathrm{C}}(\theta)}= \cos(m\theta) \vert 1 \rangle_{H,0} + \sin(m\theta) \vert 1\rangle_{V,0}.
\end{equation}
Bob measures each photon in the fixed $H/V$ linear-polarization basis, and obtains a classical outcome $x$ with possible values $H$ or $V$, distributed according to the conditional probability distribution $p_{\mathrm{G}}^{\mathrm{C}}(x\vert \theta)=|\bra{\Psi_{\mathrm{G}}^{\mathrm{C}}(\theta)}\ket{1}_{x,0}|^2$. From this, the minimum uncertainty  he can get is again dictated  by the Cram\'{e}r-Rao (CR) bound $\Delta \theta_{\mathrm{G}}^{\mathrm{C}} \geq (\nu N f)^{-1/2}$, where $f$ is the classical Fisher information per particle associated to the particular measurement strategy$^{1}$, defined as
\begin{equation}
\label{eq:CFI_def}
f=f\big(p_{\mathrm{G}}^{\mathrm{C}}(x\vert \theta)\big)= \sum_{x=H, V} \frac{1}{p_{\mathrm{G}}^{\mathrm{C}}(x \vert \theta)} \left( \frac{\partial p_{\mathrm{G}}^{\mathrm{C}}(x\vert \theta)}{\partial \theta} \right)^{2}.
\end{equation}
Substituting $p_{\mathrm{G}}^{\mathrm{C}}(H \vert \theta) =\cos^{2}(m\theta)$ and $p_{\mathrm{G}}^{\mathrm{C}}(V \vert \theta) =\sin^{2}(m\theta)$ into \eqref{eq:CFI_def}, one obtains $f=(2m)^2$. This, in turn, with the CR bound above, directly leads to optimal quantum bound (\ref{eq:QCR}). In addition, for a suitable estimation function (see Supplementary Note 2), the bound is always saturated in the limit of $\nu\to\infty$. Hence, the adopted measurement strategy allows Bob and Alice to reveal as much information about $\theta$ as allowed by quantum mechanics.

In the presence of experimental imperfections, including photon losses and non-unitary efficiencies for the $q$-plates [see Eq. (\ref{eq:state_complete})], the CR bound with polarization measurements is straightforwardly calculated to be:
\begin{equation}
\label{eq:QCR_expt}
\Delta \theta^{m}\geq \frac{1}{2 m V \sqrt{\eta} \sqrt{\nu N} \sqrt{C(\theta)}},
\end{equation}
where $C(\theta)$ is the following function:
\begin{equation}
C(\theta) = \frac{\sin(2m\theta)^2}{1-V^2 \cos(2m\theta)^2}.
\end{equation}
Hence, the minimum statistical error exhibits a dependence on the actual value of the angle, and the QCR bound can only be saturated for $\tilde{\theta}=\pi/(4m)+k \pi/(2m)$, with integer $k$. However, in Supplementary Note 2, we discuss an adaptive protocol that succeeds in saturating the QCR bound for all values of $\theta$.

%%%%%%%%%%%%%%%%%%%%%%%%%%%%%%%%%%%%%%%%%%%%%%%%%%%%%%%%%%%%%%%%%%%%%%%%%%%%%%%%%%%%%%%%%%%%%%%%%%%%%%%%%%%%%%%%%%%%%%%%%%%%%%%%%%%%%%%%%%%%%%%%%
\par
\subsection{Coherent states} 
\label{SICoherent}
Here, we show that the estimation protocol works equally well for coherent pulses of average photon number $N$ as for $N$ single photons. That is, for the same measurement strategy, the resulting precision bound is given by \eqref{eq:QCR}. We discuss explicitly only the ideal case, the case with experimental imperfections follows straightforwardly and gives (\ref{eq:QCR_expt}). Consider  a coherent state $\ket{\text{coh}(\alpha)}_{H,0}$ of generic complex amplitude $\alpha$ as input to Alice's $q$-plate.  The mean photon number is $|\alpha|^2$, and the state has $H$ linear polarization and is prepared in the zero-OAM mode. We can write $\ket{\text{coh}(\alpha)}_{H,0}=D_{H,0}(\alpha)\ket{0}$, where $D_{H,0}(\alpha)= e^{\alpha a^{\dag}_{H,0} - \alpha^{\ast} a_{H,0}}$ is the displacement operator with amplitude $\alpha$,  $a^{\dag}_{H,0}$ and $a_{H,0}$ are the creation and annihilation operators of a photon in the corresponding mode, respectively, and $\ket{0}$ represents the vacuum state. Alice's $q$-plate and HWP transform the mode operators as $a^{\dag}_{H,0}=\frac{1}{\sqrt{2}}(a^{\dag}_{R,0}+a^{\dag}_{L,0})\to\frac{1}{\sqrt{2}}(a^{\dag}_{R,-2q}+a^{\dag}_{L,2q})$, where $a^{\dag}_{R,-2q}$ ($a^{\dag}_{L,2q}$) is the creation operator of a photon in the total angular momentum eigenmode of $R$ ($L$) circular polarization and OAM $-2q$ ($2q$). The operator $a_{H,0}$ undergoes an analogous transformation. Then, using linearity and  some simple algebra, one obtains that Alice's station induces the state transformation:
\begin{equation}
\label{cohtransform}
\ket{\text{coh}(\alpha)}_{H,0}=D_{H,0}(\alpha) \vert 0 \rangle \rightarrow D_{R,-2q}\big(\alpha/\sqrt{2}\big) D_{L,2q}\big(\alpha/\sqrt{2}\big) \vert 0 \rangle =\ket{\text{coh}\big(\alpha/\sqrt{2}\big)}_{R,-2q}\ket{\text{coh}\big(\alpha/\sqrt{2}\big)}_{L,2q} = \ket{\Psi_{\mathrm{G}}^{\text{coh}}},
\end{equation}
where the last equality defines our coherent-state hybrid SAM+OAM states. 

Under rotation, $\ket{\Psi_{\mathrm{G}}^{\text{coh}}}$ evolves into $\ket{\Psi_{\mathrm{G}}^{\text{coh}}(\theta)}=\mathcal{U}_{\theta}\ket{\Psi_{\mathrm{G}}^{\text{coh}}}$, with $\mathcal{U}_{\theta} = e^{-\imath \mathcal{G} \theta}$, and $\mathcal{G}$ the generator of state rotations defined in Supplementary Note 1. Using that $\mathcal{U}_{\theta}a^{\dag}_{L/V,\pm2q}\mathcal{U}_{\theta}^{\dag}=e^{\mp im\theta}a^{\dag}_{L/V,\pm2q}$, and analogously for $a_{L/V,\pm2q}$, we have 
\begin{widetext}
\begin{eqnarray}
\label{cohexplicit}
\ket{\Psi_{\mathrm{G}}^{\text{coh}}(\theta)}=\ket{\text{coh}\big(e^{ im\theta}\alpha/\sqrt{2}\big)}_{R,-2q}\ket{\text{coh}\big(e^{- im\theta}\alpha/\sqrt{2}\big)}_{L,2q},
\end{eqnarray}
\end{widetext}
where as before $m=2q+1$. 
The quantum Fisher information $F_{N}\big(\ket{\Psi^G_{\text{coh}}(\theta)}\big)$ calculated from the probe state $\ket{\Psi_{\mathrm{G}}^{\text{coh}}(\theta)}$ is  $F_{N}\big(\ket{\Psi_{\mathrm{G}}^{\text{coh}}(\theta)}\big)=4 \langle \Psi_{\mathrm{G}}^{\text{coh}}(\theta) \vert (\Delta G)^2 \vert \Psi^G_{\text{coh}}(\theta) \rangle$, leading to the precision 
\begin{equation}
\label{eq:QCR_coh}
\Delta \theta_{\mathrm{G}}^{\text{coh}} \geq \frac{1}{2m \sqrt{\nu |\alpha|^2}}.
\end{equation}
For the particular case when the mean photon number $|\alpha|^2$ is $N$, bound \eqref{eq:QCR_coh} is equal to \eqref{eq:QCR}.

Next, we show that Bob's polarization measurements are optimal, even for the case of coherent states. To the coherent-pulse, he applies the same decoding transformations as in the single photon strategy:  a HWP followed by $q$-plate of charge $q$. The output state is
\begin{widetext}
\begin{equation}
\label{Bobbasiscoh}
\ket{\Psi^G_{\text{coh}}(\theta)} \rightarrow\ket{\text{coh}\big(e^{ im\theta}\alpha/\sqrt{2}\big)}_{R,0}\ket{\text{coh}\big(e^{- im\theta}\alpha/\sqrt{2}\big)}_{L,0}=\ket{\text{coh}\big(\cos(m\theta)\alpha\big)}_{H,0}\ket{\text{coh}\big(\sin(m\theta)\alpha\big)}_{V,0},
\end{equation}
\end{widetext}
As in the single-photon case, he sends output state through a $H/V$ polarizing beam splitter, and detects the conditional probability distribution $p(n_{H},n_{V} \vert \theta)$ of obtaining $n_{H}$ horizontally polarized photons and $n_{V}$ vertically polarized ones with an intensity measurement. The associated classical Fisher information is
\begin{widetext}
\begin{eqnarray}
\label{fishercoh}
f\big(p(n_{H},n_{V}\vert \theta)\big)&=& \sum_{n_{H},n_{V}=0}^{\infty} \frac{1}{p(n_{H},n_{V} \vert \theta)} \left( \frac{\partial p(n_{H},n_{V} \vert \theta)}{\partial \theta} \right)^{2}\text{, with}\\
\label{epform}
p(n_{H},n_{V} \vert \theta)&=&e^{-|\alpha|^2}\frac{\big(\cos^2(m\theta)|\alpha|^2\big)^{n_{H}}\big(\sin^2(m\theta)|\alpha|^2\big)^{n_{V}}}{n_{H}!n_V!},
\end{eqnarray}
\end{widetext}
where the explicit form \eqref{epform} of distribution $p(n_{H},n_{V} \vert \theta)$ is obtained by expanding the right-hand side of \eqref{Bobbasiscoh} in the Fock-state basis. A straightforward calculation shows that the CR bound applied to Fisher information \eqref{fishercoh} coincides exactly with the optimal quantum scaling \eqref{eq:QCR_coh}.  Also, as usual, with a suitable estimator, the bound is saturated in the limit $\nu\to\infty$. Hence, the polarization analysis adopted for single-photon probes is also the optimal measurement strategy for coherent-pulses.

%%%%%%%%%%%%%%%%%%%%%%%%%%%%%%%%%%%%%%%%%%%%%%%%%%%%%%%%%%%%%%%%%%%%%%%%%%%%%%%%%%%%%%%%%%%%%%%%%%%%%%%%%%%%%%%%%%%%%%%%%%%%%%%%%%%%%%%%%%%%%%%%%%%%

\section*{Supplementary Note 2: The estimation procedure}
\label{Appestimation}
Here we describe the technical details of our phase estimation procedure.
%%%%%%%%%%%%%%%%%%%%%%%%%%%%%%%%%%%%%%%%%%%%%%%%%%%%%%%%%%%%%%%%%%%%%%%%%%%%%%%%%%%%%%%%%%%%%%%%%%%%%%%%%%%%%%%%%%%%%%%%%%%%%%%%%%%%%%%%%%%%%%%%%
\subsection*{A. Bayesian estimator}
\label{Bayesian}
With the measurement strategy established, an estimator must be used to process the experimental data obtained. The experimental outcomes of $M=\nu \times N$ independent photons can be represented by the string $X=(x_{1}, \ldots, x_{M})$, with $x_i=H$ or $V$, for $1\leq i\leq M$. An estimator is a function that maps $X$ into an estimate $\overline{\theta}$ of the actual value $\theta^*$. A suitable choice is given by the Bayesian estimator$^{39}$, which is based on Bayes' rule: 
\begin{equation}
\label{Bayes}
P(\theta \vert X) P(X) = P(X \vert \theta) P(\theta).
\end{equation}
$P(\theta \vert X)$ is the conditional probability of $\theta^*$ being equal to $\theta$ given that the observed outcome string is $X$. This is the desired distribution, as from it both the estimate and its uncertainty can be directly obtained. $P(X \vert \theta)$ gives the conditional probability of getting  $X$ given that the phase is $\theta$. Since the photons are independent,  it can be decomposed as
\begin{equation}
\label{facto}
P(X \vert \theta) = \prod_{k=1}^{M} p(x_{k} \vert \theta).
\end{equation}
In addition, since the experimental setup is well-characterized, the distributions $p(x_{k} \vert \theta)$ with which one can explicitly evaluate \eqref{facto} are known. For instance, for the SAM+OAM states, these are given by 
\begin{eqnarray}
\label{pofx}
p(x_{k} \vert \theta)=p_{\mathrm{G}}^{\mathrm{C}}(x_{k}\vert \theta)=\left\{ \begin{array}{ll} 
\cos^2(m\theta) & \textrm{if $x_{k}=H$,}\\ 
\sin^2(m\theta)& \textrm{if $x_{k}=V$,}\end{array} \right. 
\end{eqnarray} 
for all $1\leq k\leq M$. In turn, $P(X)$ is the probability of getting $X$ regardless of the value of $\theta$. This can also be explicitly evaluated, as it is defined in terms of  $P(X \vert \theta)$: $P(X)=\int_{\Omega} d \theta P(X \vert \theta)$, where $\Omega$ is the interval of bijectivity of $P(X \vert \theta)$ as a function of $\theta$, of length $0<\mathcal{T}\leq 2\pi$, in which $\theta^*$ is known to lie. Finally, $P(\theta)$ is the probability of the phase being $\theta$ regardless of the detected outcomes, which is of course unknown. In Bayesian estimation, one simply guesses $\theta^*$ based on whatever apriori knowledge. That is, one substitute $P(\theta)$ by $P_{\text{prior}}(\theta)$, which describes this knowledge. When no prior knowledge about $\theta^*$ is available, except  that it belongs to $\Omega$, one has $P_{\text{prior}}(\theta)=1/\mathcal{T}$. This, together with rule \eqref{Bayes} and decomposition \eqref{facto}, gives the desired {\it a-posteriori} distribution:
\begin{equation}
\label{aposteriori}
P(\theta \vert X) = \frac{1}{\mathcal{T}P(X)} \prod_{k=1}^{M} p(x_{k} \vert \theta).
\end{equation}

\par With this, the estimate and its associated mean square statistical error are obtained as
\begin{eqnarray}
\label{estimate}
\overline{\theta} &=& \int_{\Omega} d\theta \, \theta P(\theta \vert X), \\
\label{uncertainty}
\text{and } \Delta\theta^2 &=& \int_{\Omega} d\theta \, (\theta - \overline{\theta})^{2} P(\theta \vert X),
\end{eqnarray}
respectively. This estimation displays the following two convenient properties$^{40}$: (i) It is {\it locally unbiased}. This means that, in the asymptotic limit of large $M$, $\overline{\theta}$ converges to the true value $\theta^{\ast}$, for arbitrary $\theta^*$ in $\Omega$. (ii) The Bayesian estimator displays phase-independent sensitivity. That is, in the same asymptotic limit, it saturates the Cram\'{e}r-Rao inequality, for any $\theta^*\in\Omega$. Furthermore, it has been shown$^{39,41}$ that the output phase distribution obtained with Bayesian analysis reaches a Gaussian function faster than other approaches, such as Maximum-likelihood estimators, leading to a faster convergence when the sample size is small.

%%%%%%%%%%%%%%%%%%%%%%%%%%%%%%%%%%%%%%%%%%%%%%%%%%%%%%%%%%%%%%%%%%%%%%%%%%%%%%%%%%%%%%%%%%%%%%%%%%%%%%%%%%%%%%%%%%%%%%%%%%%%%%%%%%%%%%%%%%%%%%%%%%%%
\subsection*{B. Concatenated adaptative estimation strategy}
\label{Adaptive}
As $q$ increases, the oscillation frequency of distributions \eqref{pofx} grows, so that the length of their longest intervals of bijectivity, as functions of $\theta$, decreases with $q$. Therefore, if the Bayes estimator is directly applied to estimate a phase $\theta^*\in[0,2\pi)$, the resulting estimate \eqref{estimate} will only be defined up to a degeneracy that increases with $q$. To circumvent this, we adopt an estimation strategy with adaptive concatenated steps. We show here that the maximal number of concatenated steps needed to fully remove any ambiguity in the estimate is only 3. For simplicity, we describe the strategy in the single-photon regime, its extension to the coherent-pulse regime being trivially analogous.

We split the measurements on all $M=\nu N$ of photons into 3 different kinds, each one constituting a step of the strategy. Each step $j$, for $1\leq j \leq 3$, consumes $M_j$ photons prepared with $m=m_j$, in such a way that $m_3>m_{2}>m_1$, where $m_j=2q_j \pm 1$. This guarantees that every consecutive step features a higher sensitivity. Each step $j$ renders an estimate $\overline{\theta}_j$ and an uncertainty $\Delta\theta_j$, calculated respectively through \eqref{estimate} and \eqref{uncertainty}. Each estimate $\overline{\theta}_j$ is unambiguously defined only over an interval $\Omega_j$ of length $\mathcal{T}_j$ equal to half a period of oscillation of distributions \eqref{pofx} for $q=q_j$: 
\begin{equation}
\label{Tsubj}
\mathcal{T}_j=\frac{\pi}{2m_j}.
\end{equation}
In addition, the photons in each $j$-th step are prepared in the state $\ket{\Psi^G(\xi_j)}=\frac{1}{\sqrt{2}}(e^{i\xi_j}\vert 1\rangle_{R,-2q}+e^{-i\xi_j}\vert 1\rangle_{L,2q})$. The relative phase $\xi_j$ is such that the estimate $\overline{\theta}_{j-1}$ of the previous $(j-1)$-step sits exactly at one of the points of maximal sensitivity of $\ket{\Psi_{\mathrm{G}}^{\mathrm{C}}(\xi_j)}$. That is, for example:
\begin{equation}
\label{phasechoice}
 \xi_j=\frac{\pi}{4}- m_j\overline{\theta}_{j-1}.
\end{equation}
This condition is always possible to satisfy for $j>1$, where an estimate $\overline{\theta}_{j-1}$ is available. However, for the first step, unless one has some a-priori knowledge bout $\theta^*$, $\overline{\theta}_{1}$ is not defined. In this case we simply choose $\xi_1=0$. The aim of adapting the phase at each step is two-fold. On the one hand, since this allows us to achieve the maximal angular resolution of each step, it speeds up the asymptotic saturation of the QCR bound. On the other hand, as we discuss below, it breaks the symmetry in distributions \eqref{pofx} between $m_j$ and $m_{j-1}$. This allows us to reduce the potential ambiguities in the estimate by a factor of two per step. For this reason, as we explain next, only three concatenated steps suffice for arbitrary $\theta^*\in[0,2\pi)$.

For the strategy with phase adaptation, the outcomes of the $j$-th step are governed by the distributions $p_j(x_k|\theta)=\cos^2[m_j\theta+\xi_j]$, if $x_{k}=H$, or $p_j(x_k|\theta)=\sin^2[m_j\theta+\xi_j]$, if $x_{k}=V$. In the first step we always set $m_1=1$ ($q$-plate detuned, corresponding to polarization-only strategy), $\xi_1=0$, and $\Omega_1=[0,\pi/2)$. Since $p_1(x_k|\theta)$ repeat their values four times over $[0,2\pi)$, the estimate $\overline{\theta}_1\in\Omega_1$ of this step is four-fold degenerate. The actual value of $\theta^*\in[0,2\pi)$ may be either $\overline{\theta}_1$, $\pi-\overline{\theta}_1$, $\overline{\theta}_1+\pi$, or $2 \pi - \overline{\theta}_1$. For the second step, apart from taking $m_2>m_1=0$, with the exact charge determined below, we set $\xi_2$ in terms of the estimated $\overline{\theta}_1$, as determined by Eq. \eqref{phasechoice}, and $\Omega_2=[\overline{\theta}_1-\mathcal{T}_2/2,\overline{\theta}_1+\mathcal{T}_2/2)$, with $\mathcal{T}_2=\frac{\pi}{2m_2}$ as given by \eqref{Tsubj}. Unless $\overline{\theta}_1$ happens to be equal to $\mathcal{T}_2/2$, so that $\xi_2=\xi_1=0$, this phase choice causes two out of the four possible $\theta$'s to be consistent with one value of $p_2(x_k|\theta)$ and the other two with another value. This leads to a second estimate $\overline{\theta}_2\in\Omega_2$ with only a two-fold ambiguity. On the other hand, if $\overline{\theta}_1=\mathcal{T}_2/2$, the symmetry between $p_1(x_k|\theta)$ and $p_2(x_k|\theta)$ is not broken by this choice. However, one can simply choose $\xi_2=\xi_1+\varphi_2=\varphi_2$, for some suitable $\varphi_2$ that is not a multiple of $\mathcal{T}_2/2$. This decreases the angular resolution per probe, but allows one in return to break the symmetry. The third step is analogous to the second one: Apart from $m_3>m_2$, we set $\xi_3$ as given by \eqref{phasechoice} and $\Omega_3=[\overline{\theta}_2-\mathcal{T}_3/2,\overline{\theta}_2+\mathcal{T}_3/2)$, with $\mathcal{T}_3=\frac{\pi}{4m_3}$. Unless $\overline{\theta}_2$ happens to be equal to $\mathcal{T}_3/2$, this leads to a third estimate $\overline{\theta}_3\in\Omega_3$, without any ambiguity. Thus, arbitrary $\theta^*\in[0,2\pi)$ are unequivocally estimated by the final estimate  $\overline{\theta}_3$.

We next discuss the exact charge values $m_j=2q_j\pm1$ used in each step. It is not possible to increase the angular resolution arbitrarily much from step to step. In particular, one must require that 
\begin{equation}
\label{Taudelta}
\mathcal{T}_j\geq\Delta\theta_{j-1},
\end{equation}
for all $j$, so that the regions of bijectivity of $p_j(x_k|\theta)$ are not shorter than the precision of the $(j-1)$-th estimation. Also, we take $M_1\lesssim M_2^{1/2}\lesssim M_3^{1/4}$. This choice is convenient because, in the asymptotic limit of large $M=M_1+M_2+M_3$, it makes $M_1$ negligible with respect to $M_2$, and the latter in turn negligible with respect to $M_3$. So, $M_3$ tends to $M$, and the QCR bound is still asymptotically saturated$^{42,43}$. The first  uncertainty $\Delta\theta_1$ is such that $\Delta\theta_1\geq\big[2m_1\sqrt{M_1}\big]^{-1}=\big[2\sqrt{M_1}\big]^{-1}$. The uncertainty $\Delta\theta_2$ in the second step is such that $\Delta\theta_2\geq\big[2\sqrt{ m_1^{2}M_1+m_2^{2} M_2}\big]^{-1}$, which is approximately equal to $\big[2m_2\sqrt{M_2}\big]^{-1}$ in the limit of large $M_2$. The latter conditions for $\Delta\theta_1$ and $\Delta\theta_2$, together with \eqref{Taudelta} and \eqref{Tsubj}, lead to
\begin{equation}
2q_{j}+1\leq (2q_{j-1}+1)\pi\sqrt{M_{j-1}},
\end{equation}
for all $j$. This relationship tells us that the more we want to increase the angular resolution (the charge) from step $j-1$ to step $j$, the higher the precision of step $j-1$ must be, by increasing the number of probes  $M_{j-1}$. Or, equivalently, given fixed numbers of probes $M_{1}$, $M_{2}$, and $M_{3}$, the relationship sets a limit to the maximal jumps in resolution we can take per step. We take the optimal choice
\begin{equation}
\label{qjqj+1}
2q_{j}+1= \lfloor(2q_{j-1}+1)\pi\sqrt{M_{j-1}}\rfloor,
\end{equation}
where $\lfloor(2q_{j-1}+1)\pi\sqrt{M_{j-1}}\rfloor$ stands for the largest integer smaller than $(2q_{j-1}+1)\pi\sqrt{M_{j-1}}$. In the asymptotic limit of large $M$, the overall final uncertainty $\Delta\theta_3$ tends to the minimal uncertainty
\begin{equation}
\label{finaldelta}
\big[2\sqrt{m_1^{2} M_1+m_2^{2} M_2+m_3^{2} M_3} \big]^{-1}  \sim \big[2 m_3 \sqrt{M} \big]^{-1},
\end{equation}
and saturates the QCR bound$^{42,43}$.

Also, even though minimal uncertainty \eqref{finaldelta} is in principle independent of $\theta$, experimental imperfections make it divergent for some values of $\theta^*$. The phase-adaptive strategy described above allows one to avoid these divergencies, so that an approximate phase-independent sensitivity is recovered. Furthermore, this adaptive approach does not introduce any modification of the setup, since it requires inserting $q$-plates with different values, to be switched on and off by electrical tuning, and different linear-polarization states input.

%%%%%%%%%%%%%%%%%%%%%%%%%%%%%%%%%%%%%%%%%%%%%%%%%%%%%%%%%%%%%%%%%%%%%%%%%%%%%%%%%%%%%%%%%%%%%%%%%%%%%%%%%%%%%%%%%%%%%%%%%%%%%%%%%%%%%%%%%%%%%%%%%%%%%%%%%%%%%%

%%%%%%%%%%%%%%%%%%%%%%%%%%%%%%%%%%%%%%%%%%%%%
\section*{Supplementary References}

\begin{small}

\noindent $^{39}$ L. Pezz\`e, A. Smerzi, G. Khoury, J. F. Hodelin, and D. Bouwmeester, 
{\it Phase Detection at the Quantum Limit with Multiphoton Mach-Zehnder Interferometry}, 
Phys. Rev. Lett. {\bf 99}, 223602 (2007). 

\noindent $^{40}$ B. Teklu, S. Olivares, and M. G. A. Paris, 
{\it Bayesian estimation of one-parameter qubit gates},
J. Phys. B: At. Mol. Opt. Phys {\bf 42}, 035502 (2009). 

\noindent $^{41}$ R. Krischek, C. Schwemmer, W. Wieczorek, H. Weinfurter, P. Hyllus, L. Pezz\`e, and A. Smerzi,
{\it Useful Multiparticle Entanglement and Sub-Shot-Noise Sensitivity in Experimental Phase Estimation},
Phys. Rev. Lett. {\bf 107}, 080504 (2011).
 
\noindent $^{42}$ H. Nagaoka, 
{\it  An asymptotic efficient estimator for a one-dimensional parametric model of quantum statistical operators},
Proc. Int. Symp. on Inform. Th. 198 (1988).

\noindent $^{43}$ O. E. Barndorff and R. D. Gill, 
{\it Fisher information in quantum statistics}, 
J. Phys. A {\bf 33}, 4481-4490 (2000).

\end{small}